\documentclass[trackchanges]{aastex701}
\usepackage{amsmath}
\usepackage{nameref}
\usepackage{CJKutf8}

\begin{document}

\title{PIFFLE: Characterizing Foreground Dispersion Measure Contributions Across 4 Decades in Halo Mass in the FRB20230907D Field}

\author[]{Qi Guo}
\affiliation{Kavli IPMU (WPI), UTIAS, The University of Tokyo, Kashiwa, Chiba 277-8583, Japan}
\email[show]{2463000963@g.ecc.u-tokyo.ac.jp}  

\author[0000-0001-9299-5719]{Khee-Gan Lee}
\affiliation{Kavli IPMU (WPI), UTIAS, The University of Tokyo, Kashiwa, Chiba 277-8583, Japan}
\affiliation{Center for Data-Driven Discovery, Kavli IPMU (WPI), UTIAS, The University of Tokyo, Kashiwa, Chiba 277-8583, Japan}
\email{kglee@ipmu.jp}

\author[]{Sunil Simha}
\affiliation{Center for Interdisciplinary Exploration and Research in Astronomy, Northwestern University,1800 Sherman Avenue, Evanston, IL 60201, USA}
\affiliation{Department of Astronomy and Astrophysics, University of Chicago, William Eckhardt Research Center, 5640 S Ellis Ave, Chicago, IL 60637}
\email{sunil.simha95@gmail.com}

\author[]{Chenze Dong}
\affiliation{Kavli IPMU (WPI), UTIAS, The University of Tokyo, Kashiwa, Chiba 277-8583, Japan}
\email{chenze.dong@ipmu.jp}

\author[]{Ilya S. Khrykin}
\affiliation{Instituto de F\'isica, Pontificia   Universidad Cat\'olica de Valpara\'iso, Casilla 4059, Valpara\'iso, Chile}
\email{i.khrykin@gmail.com}

\author[]{Nicolas Tejos}
\affiliation{Instituto de F\'isica, Pontificia   Universidad Cat\'olica de Valpara\'iso, Casilla 4059, Valpara\'iso, Chile}
\email{nicolas.tejos@pucv.cl}

\author[]{J. Xavier Prochaska}
\affiliation{Department of Astronomy and Astrophysics, University of California, Santa Cruz 1156 High Street Santa Cruz CA 95064, USA }
\affiliation{Kavli IPMU (WPI), UTIAS, The University of Tokyo, Kashiwa, Chiba 277-8583, Japan}
\email{xavier@ucolick.org}

\author[]{Kevin McCarthy}
\affiliation{Kavli IPMU (WPI), UTIAS, The University of Tokyo, Kashiwa, Chiba 277-8583, Japan}
\email{kevin.mccarthy@ipmu.jp}

\author[]{John D. Silverman}
\affiliation{Kavli IPMU (WPI), UTIAS, The University of Tokyo, Kashiwa, Chiba 277-8583, Japan}
\email{john.silverman@ipmu.jp}

\author[]{Ines Pastor-Marazuela}
\affiliation{ASTRON, Netherlands Institute for Radio Astronomy, Oude Hoogeveensedijk 4, 7991 PD Dwingeloo, the Netherlands}
\affiliation{Anton Pannekoek Institute for Astronomy, University of Amsterdam, Science Park 904, 1098 XH, Amsterdam, the Netherlands}
\email{ines.pastor-marazuela@manchester.ac.uk}

\author[]{Ben Stappers}
\affiliation{Jodrell Bank Centre for Astrophysics, University of Manchester, Oxford Road, Manchester M13 9PL, UK}
\email{Ben.Stappers@manchester.ac.uk}

\author[]{K. M. Rajwade}
\affiliation{Astrophysics, University of Oxford, Denys Wilkinson Building, Keble Road, Oxford OX1 3RH, UK
}
\email{kaustubh.rajwade@dtc.ox.ac.uk}

\author[]{Manisha Caleb}
\affiliation{Sydney Institute for Astronomy, School of Physics, The University of Sydney, Sydney, Australia}
\affiliation{ARC Centre of Excellence for Gravitational Wave Discovery (OzGrav), Australia
}
\email{manishacaleb@gmail.com}

\author{Bonny Y.~Wang ({\begin{CJK*}{UTF8}{bsmi}汪玥\end{CJK*}})}
\affiliation{Department of Astronomy and Astrophysics, University of Chicago, William Eckhardt Research Center, 5640 S Ellis Ave, Chicago, IL 60637}
\affiliation{Center for Data-Driven Discovery, Kavli IPMU (WPI), UTIAS, The University of Tokyo, Kashiwa, Chiba 277-8583, Japan}
\affiliation{Kavli IPMU (WPI), UTIAS, The University of Tokyo, Kashiwa, Chiba 277-8583, Japan}
\affiliation{NSF-Simons AI Institute for the Sky (SkAI), 172 E. Chestnut St., Chicago, IL 60611, USA}
\email{bonnyw@uchicago.edu}

\author{Leander Thiele}
\affiliation{Kavli IPMU (WPI), UTIAS, The University of Tokyo, Kashiwa, Chiba 277-8583, Japan}
\affiliation{Center for Data-Driven Discovery, Kavli IPMU (WPI), UTIAS, The University of Tokyo, Kashiwa, Chiba 277-8583, Japan}
\email{leander.thiele@ipmu.jp}

\author[0000-0001-6869-2996]{Zhao Joseph Zhang
({\begin{CJK*}{UTF8}{gbsn}张钊\end{CJK*}})}

\affiliation{Theoretical Astrophysics, Department of Earth and Space Science,
The University of Osaka, 1-1 Machikaneyama, Toyonaka,
Osaka 560-0043, Japan}

\affiliation{Department of Physics, The University of Hong Kong,
Pokfulam Road, Hong Kong, China}

\email{zzhang@astro-osaka.jp}

\author[0000-0001-7457-8487]{Kentaro Nagamine}
\affiliation{Theoretical Astrophysics, Department of Earth and Space Science, Graduate School of Science, The University of Osaka, 1-1 Machikaneyama, Toyonaka, Osaka, 560-0043, Japan}
\affiliation{Theoretical Joint Research, Forefront Research Center, The University of Osaka, 1-1 Machikaneyama, Toyonaka, Osaka 560-0043, Japan}
\affiliation{Kavli IPMU (WPI), UTIAS, The University of Tokyo, Kashiwa, Chiba 277-8583, Japan}
\affiliation{Department of Physics and Astronomy, University of Nevada, Las Vegas, 4505 S. Maryland Pkwy, Las Vegas, NV 89154-4002, USA}
\affiliation{Nevada Center for Astrophysics, University of Nevada, Las Vegas, 4505 S. Maryland Pkwy, Las Vegas, NV 89154-4002, USA}
\email{kn@astro-osaka.jp}

\begin{abstract}
We characterize the foreground environment of FRB20230907D, localized to a galaxy at $z=0.464$, which has an observed dispersion measure of ${\rm DM}_{\rm obs}=1031~{\rm pc~cm^{-3}}$.
At its redshift, FRB20230907D lies above the Macquart relation, the expected relation between cosmological dispersion measure and the source redshift, indicating a substantial excess DM along this line of sight.
We use Subaru/PFS and SDSS spectroscopy, published group catalogs, Rubin/LSST imaging, and eROSITA X-ray data to characterize the foreground structures that may account for this excess.
A friends-of-friends search identifies a massive foreground system at $z\simeq0.09$ with $M_{\rm vir}\simeq7.7\times10^{14}~M_\odot$, while low redshift catalogs reveal an additional group at $z\simeq0.02565$.
Assuming that the halo gas follows a modified-NFW halo density profile, we estimate observer frame contributions of $205^{+30}_{-25}~{\rm pc~cm^{-3}}$ and $80^{+60}_{-40}~{\rm pc~cm^{-3}}$ from these systems, respectively.
After accounting for the remaining foreground and host contributions, we are able to make the first-ever constraints on the dispersion measure of a closely-intersected $M_*\approx 10^8\,M_\odot$ dwarf galaxy: ${\rm DM}_{\rm UGC7596}=58^{+53}_{-40}~{\rm pc~cm^{-3}}$ at $b_\perp \approx 0.03\,R_{\rm vir}$, which is consistent with hydrodynamical predictions for low mass halos.
This highlights the enhancement of dense foreground spectroscopy and multi-wavelength data toward the interpretation of FRB dispersion measures.
\end{abstract}



\section{Introduction}
\label{sec:intro}

One of the goals of observational cosmology is to understand how baryonic matter is distributed in the universe.
The cosmic baryon density at the early Universe is tightly constrained by Big Bang nucleosynthesis (BBN) \citep{Burles2001,Cyburt2016} and the cosmic microwave background (CMB) measurements, i.e. $\Omega_{\rm b}h^2\sim0.02$ \citep{Planck2018}.
At low redshift ($z\lesssim2$), however, direct inventories of stars, cold interstellar gas, and the hot intracluster medium account for approximately $35\%$ of the cosmic baryon budget inferred from these early Universe measurements \citep{Fukugita2004}, leading to the so-called "missing baryon problem" \citep{Fukugita1998,Shull2012}.
A substantial fraction of the remaining baryons is expected to reside in diffuse, ionized gas distributed throughout the intergalactic medium (IGM) and/or the circumgalactic medium (CGM) of galaxies, groups, and clusters \citep{Cen1999,Cen2006,Prochaska2019,Macquart2020,Khrykin2024b}.
Because this diffuse ionized gas is typically low density, with a substantial fraction expected to reside at warm hot temperatures ($T\sim10^{5}$--$10^{7}\,{\rm K}$), its emission is weak and determining its spatial distribution remains observationally challenging.

At $z\gtrsim2$, the photoionized IGM can be traced through absorption features such as the H\,{\sc I} Lyman-$\alpha$ forest in the spectra of background quasars \citep{Rauch1997,Weinberg1997,Meiksin2009,McQuinn2016}.
With sufficiently dense samples of background sightlines, these absorption measurements can also be used to reconstruct the three-dimensional large scale structure of the IGM through tomographic techniques \citep{Pichon2001,Caucci2008,Lee2014,Lee2016,Lee2018}.
However, absorption line measurements depend on the ionization state and, for metal lines, the metallicity of the gas.
Since much of the diffuse baryonic matter in the IGM and the CGM is ionized, its free electrons provide a direct tracer of this baryonic component.
An independent probe that is directly sensitive to the free electron content of the Universe is therefore valuable.

Fast radio bursts (FRBs) provide such a probe.
FRBs are short duration radio transients characterized by millisecond-timescale pulses and extremely high inferred brightness temperatures \citep{Lorimer2007,Thornton2013,Petroff2019,Cordes2019}.
One of their defining observational signatures is a frequency dependent arrival time delay caused by the propagation of radio waves through ionized plasma, with $\Delta t\propto {\rm DM}\,\nu^{-2}$, where DM denotes the dispersion measure.
The observed DM is the redshift ($z$) weighted free electron column density along the line of sight,
\begin{equation}
{\rm DM}_{\rm obs}
=
\int_0^{z_{\rm FRB}}
\frac{n_e(z)}{1+z}
\,{\rm d}l_{\rm prop},
\end{equation}
where $n_e(z)$ is the physical free electron number density and ${\rm d}l_{\rm prop}$ is the proper line element
(\citealt{Lorimer2004},\citealt{Ioka2003},\citealt{Inoue2004}; see \citealt{Glowacki2026} for a review).
Because DM depends directly on the integrated free electron density, it traces ionized gas without requiring assumptions about gas metallicity or the ionization fraction of any particular element.

The rapid development of FRB observing facilities has greatly increased both the number of detected bursts and the number of accurately localized events.
The Canadian Hydrogen Intensity Mapping Experiment (CHIME) has provided large FRB samples for population studies \citep{CHIME2021}, while interferometric facilities such as the Australian Square Kilometre Array Pathfinder (ASKAP) and the Deep Synoptic Array (DSA) provide the precise sky localizations required to identify host galaxies and measure source redshifts \citep{Shannon2018,Bannister2019,Macquart2020,Kocz2019,Ravi2023}.
MeerKAT is particularly powerful for FRB discovery and localization because of its high sensitivity and interferometric capabilities.
The commensal MeerTRAP project performs real-time FRB searches with MeerKAT, while its transient-buffer system enables arcsecond and sub-arcsecond localizations \citep{Rajwade2024,Jankowski2023}.
Its sensitivity also enables localization of relatively high redshift FRBs, making MeerKAT sightlines especially valuable for cosmological and foreground environment studies.
Other facilities, including FAST and the VLA/realfast system, provide complementary sensitivity, frequency coverage, and localization capabilities \citep{Nan2011,Law2018,Niu2022}.

In terms of using FRBs as a cosmological probe, for an extragalactic FRB, the observed DM can be decomposed as
\begin{equation}
{\rm DM}_{\rm obs}
=
{\rm DM}_{\rm MW}
+
{\rm DM}_{\rm cosmic}
+
\frac{{\rm DM}_{\rm host}}{1+z_{\rm FRB}},
\end{equation}
where ${\rm DM}_{\rm MW}$ is the contribution from the MW interstellar medium (ISM) and halo, ${\rm DM}_{\rm host}$ is the rest frame contribution from the host galaxy and the local environment of the source, and ${\rm DM}_{\rm cosmic}$ is the contribution accumulated between the MW and the host galaxy \citep{Ioka2003,Inoue2004,Macquart2020}.
The cosmic term may be further divided into
\begin{equation}
{\rm DM}_{\rm cosmic}
=
{\rm DM}_{\rm IGM}
+
{\rm DM}_{\rm halos},
\end{equation}
where ${\rm DM}_{\rm IGM}$ represents diffuse gas associated with the large scale cosmic web tracing matter overdensities of $\delta_m \sim 0-10$ while ${\rm DM}_{\rm halos}$ represents gas associated with the halos of intervening galaxies, groups, and clusters \citep{Prochaska2019,Simha2020,Lee2022}.

The observed relation between ${\rm DM}_{\rm cosmic}$ and redshift for localized FRBs, commonly referred to as the Macquart relation, is broadly consistent with the baryon density inferred from early Universe measurements \citep{Macquart2020}.
At fixed redshift, individual sightlines intersect different combinations of voids, sheets, filaments, and halos, giving rise to an intrinsic sightline-to-sightline dispersion in ${\rm DM}_{\rm cosmic}$\citep{McQuinn2014,Macquart2020,Batten2021,Batten2022}.
Galaxy formation feedback can modify the width and shape of this distribution by redistributing baryons between halos and the diffuse IGM \citep{Medlock2024,Guo_2025,Caleb2026}.
The mean relation, however, is much less sensitive to these details than the dispersion about it.
This sightline-to-sightline variation is not only a source of uncertainty, but also contains information about the spatial distribution of baryons.

One approach to decode this DM variance is by combining localized FRBs with foreground galaxy surveys to identify the large scale structures and intervening halos responsible \citep{Simha2020,Lee2022}.
In this foreground mapping approach, spectroscopic redshifts are used to reconstruct the three-dimensional galaxy distribution along an FRB sightline.
The diffuse IGM contribution can then be modeled from the reconstructed density field, while the halo contribution can be estimated from the properties of individual foreground galaxies, groups, and clusters \citep{Lee2022,Lee2023,Khrykin2024b}.
Unlike an analysis based only on the statistical DM--redshift relation, this approach uses information from the actual foreground of each FRB as conditioning information to study the diffuse cosmic baryons.

Foreground spectroscopy is particularly important for FRBs whose DMs differ substantially from the mean Macquart relation.
Without detailed information about the line of sight, an apparent DM excess may be attributed to the diffuse IGM, the host environment, or one or more intervening halos.
These possibilities can be partially separated by identifying foreground structures and measuring their redshifts, projected separations, velocity distributions, and halo masses \citep{Lee2022,Khrykin2024b}.
Dense spectroscopy therefore provides the spatial information needed to associate portions of the integrated electron column with specific structures along the line of sight.

Galaxy groups and clusters are potentially important contributors to ${\rm DM}_{\rm halos}$ because their extended ionized gas reservoirs can produce substantial DMs over large projected areas. E.g., Clusters along the sightline of FRB 20190520B could explain the very large excess in DM, a source whose host contribution to the DM budget was initially thought to be significantly large \cite{Lee2023}. 
Identifying these systems from redshift survey data, however, is not trivial.
Friends-of-friends group finders depend on the adopted transverse and line-of-sight linking lengths, as well as on the magnitude limit, spatial sampling, and spectroscopic completeness of the survey.
Inappropriate linking parameters can fragment a single halo into several recovered systems or merge physically distinct structures into one group.
Group memberships and dynamical mass estimates should therefore be tested using mock catalogs in which the underlying halo associations are known.

FRB20230907D provides a sightline for such an analysis.
Using the foreground corrections adopted by \citet{PastorMarazuela2025}, the inferred Macquart relation redshift is $z_{\rm Macquart}=0.84^{+0.18}_{-0.44}$, higher than the spectroscopic host galaxy redshift, $z_{\rm FRB}=0.464$.
This discrepancy indicates that the observed DM is high relative to the mean Macquart relation at the measured source redshift and motivates a more complete reconstruction of the foreground along this sightline.

In this work, we use dense wide field spectroscopy from the \textquoteleft{}\={O}nohi\textquoteleft{}ula Subaru Prime Focus Spectrograph (PFS), supplemented by archival spectroscopy and multiwavelength data, to reconstruct the foreground of FRB20230907D.
The sightline passes through the projected Virgo--M49 environment, close to the dwarf galaxy UGC~7596, and intersects a prominent cluster scale foreground structure near $z\simeq0.09$.
We identify the foreground structures relevant to the sightline and estimate their contributions to the observed DM.
The remainder of this paper is organized as follows.
Section~\ref{sec:method} describes the observational data and group finding methodology.
Section~\ref{sec:results} presents the mock catalog validation and the resulting foreground group catalog.
Section~\ref{sec:dm_inference} presents the DM inference for the identified foreground structures and the other components along the sightline.
Finally, Section~\ref{sec:conclusion} summarizes our main results.

\section{Data and Methodology}\label{sec:method}
\subsection{FRB20230907D field}

FRB20230907D is discovered by MeerKAT \citep{PastorMarazuela2025}.
It is located at 
${\rm RA}=12^{\rm h}28^{\rm m}34.20^{\rm s}\pm0.40\arcsec$ and ${\rm Dec}=+08^\circ39'29.13''\pm0.57\arcsec$, with an observed ${\rm DM}_{\rm obs}=1030.79\pm0.04~{\rm pc~cm^{-3}}$.
It is localized to a galaxy at $z_{\rm FRB}=0.464 \pm (9.533\times10^{-6})$ using Keck/LRIS spectroscopy.
We adopt $z_{\rm FRB}=0.464$ as the source redshift in this work.

\begin{figure}
\centering
\includegraphics[width=0.95\linewidth]
{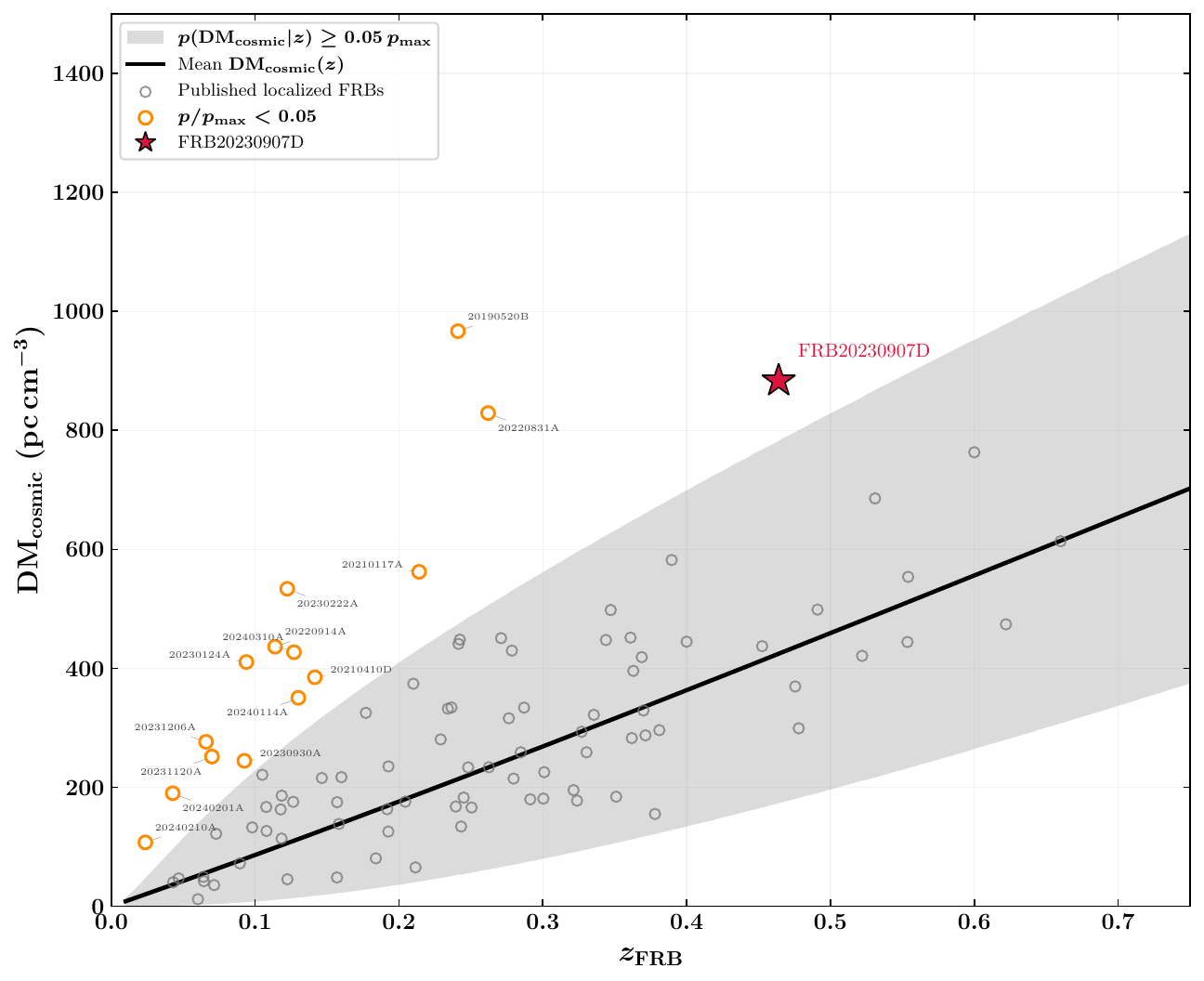}
\caption{
FRB20230907D compared with the ${\rm DM}_{\rm cosmic}$--$z$ distribution computed using the FRB covariance framework of \citet{Reischke2023}.
The black curve shows the mean cosmological DM predicted by the model, and the gray region shows where the conditional probability density at fixed redshift satisfies $p({\rm DM}_{\rm cosmic}|z)\geq0.05\,p_{\rm max}$, where $p_{\rm max}(z)\equiv\max_{{\rm DM}_{\rm cosmic}}p({\rm DM}_{\rm cosmic}|z)$ is the maximum value of the conditional probability density at that redshift. 
Gray circles show localized FRBs from the compilation of \citet{ReischkeHagstotz2026}, while orange circles indicate sources lying outside this displayed high probability region under the adopted MW and host DM corrections.
Orange labels omit the ``FRB'' prefix for clarity.
The red star marks FRB20230907D at $z_{\rm FRB}=0.464$.
}
\label{fig:frb20230907_macquart}
\end{figure}

Figure~\ref{fig:frb20230907_macquart} compares FRB20230907D with the ${\rm DM}_{\rm cosmic}$--$z$ distribution computed using the FRB covariance framework of \citet{Reischke2023}.
For context, we also show the localized FRB compilation of \citet{ReischkeHagstotz2026}, restricted to the redshift range shown in the figure.
For each FRB, we estimate ${\rm DM}_{\rm cosmic}$ by subtracting the MW ISM and halo contributions and a representative rest frame host contribution of ${\rm DM}_{\rm host,rest}=100~{\rm pc~cm^{-3}}$ from the observed DM.
For the comparison sample, we subtract the catalogued MW ISM contribution and adopt an additional ${\rm DM}_{\rm MW,halo}=50~{\rm pc~cm^{-3}}$ \citep{Prochaska2019}, which is fixed only for the illustrative comparison.
For FRB20230907D, the NE2001 model \citep{Cordes2004} gives ${\rm DM}_{\rm MW,ISM}=29~{\rm pc~cm^{-3}}$ \citep{PastorMarazuela2025}.
Including the same adopted MW contribution gives a total ${\rm DM}_{\rm MW}=79~{\rm pc~cm^{-3}}$.
FRB20230907D is thus seen to lie well above the mean cosmological DM expected at its measured redshift.

\begin{figure}
\centering
\includegraphics[width=0.95\textwidth]{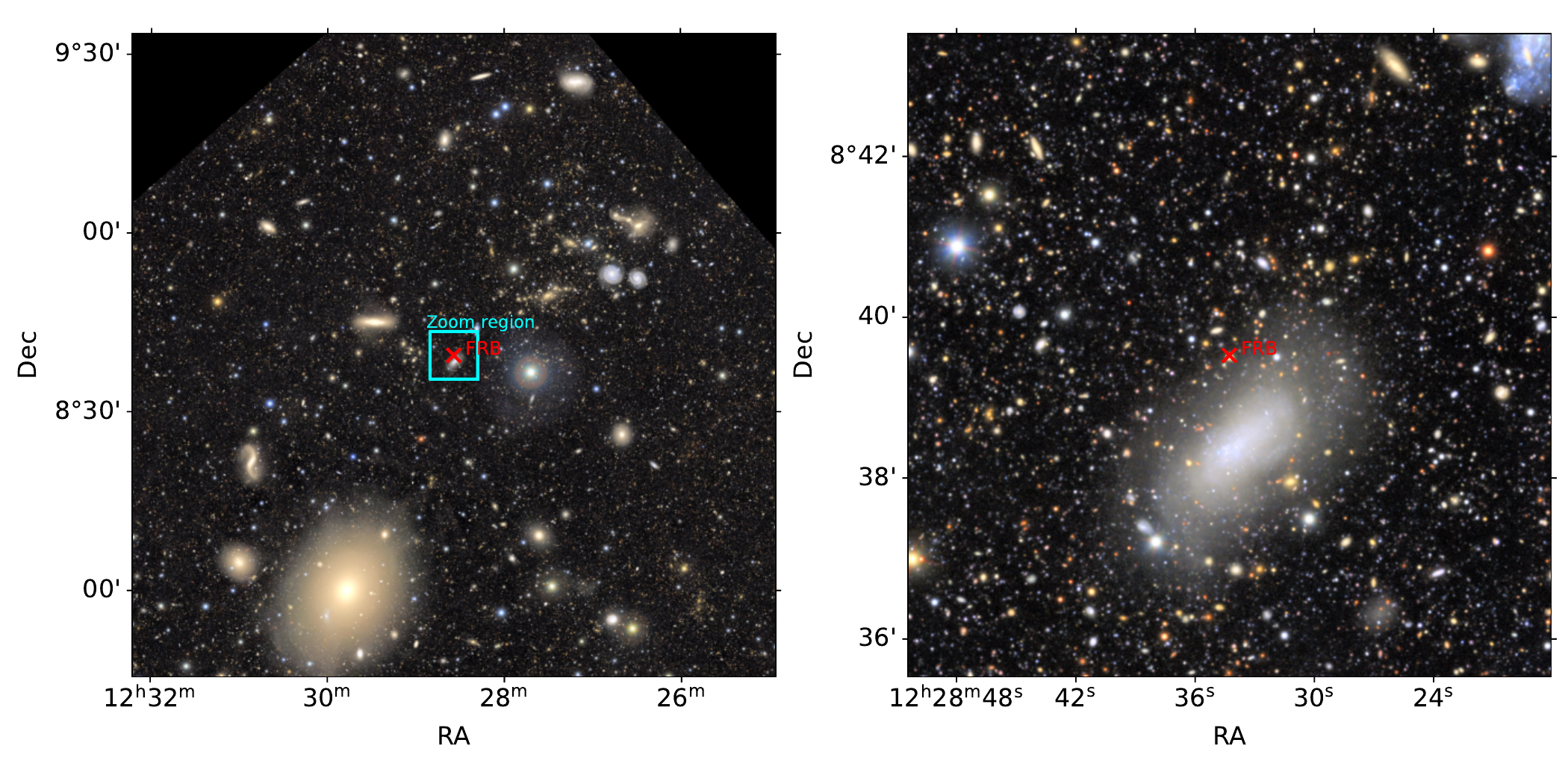}
\caption{
LSST/Rubin First Look image \citep{Ivezic2019,RubinFirstLook2025} of the FRB20230907D field.
The left panel shows the wider field around the FRB, while the right panel zooms in on the region close to the burst position.
The red cross marks the FRB position.
The FRB lies close in projection to the dwarf galaxy UGC~7596.
}
\label{fig:frb20230907_lsst}
\end{figure}

The field is shown in Figure~\ref{fig:frb20230907_lsst}, using optical imaging from the Rubin Observatory/LSST First Look data \citep{Ivezic2019,RubinFirstLook2025}.
The FRB position lies close in projection to the dwarf galaxy UGC~7596, making this sightline immediately interesting for foreground studies.

Given that the Rubin/LSST image reveals that FRB20230907D intersects the extended stellar component of UGC 7596, it is worth revisiting the possibility that UGC 7596 is in fact the host. Given its proximity (D=7.78 Mpc \citep{Daz-Garca2016}), this would imply that intervening DM contributions are negligible and nearly all the measured $\mathrm{DM}\approx 1000\,{\rm pc~cm^{-3}}$ has to be assigned to the host.
This would surpass by far all known FRBs, even the restframe $\mathrm{DM_{host}}\approx 700\,{\rm pc~cm^{-3}}$ originally reported for FRB 20190520B \citep{Niu2022} --- although this is  now shown to be spurious \citep{Lee2023}.

\citet{PastorMarazuela2025} originally identified their source G1 at $z=0.464$ as the putative host using the PATH framework \citep{Aggarwal2021}, obtaining $P(O|x)=94.2\%$ among the three candidate galaxies considered without incorporating information about the FRB DM.
As a separate test between G1 and UGC~7596, we additionally incorporate the DM-weighted host association analysis using the $z$-DM framework of \citet{James2022b}, adopting the MeerTRAP coherent survey configuration \citep{Jankowski2023}.
We obtain $P(O_{\rm G1}|x,{\rm DM}_{\rm obs}) = 99.998 \%$ and $P(O_{\rm UGC7596}|x,{\rm DM}_{\rm obs}) = 0.002 \%$, strongly disfavoring UGC~7596 as the host under the adopted model.

\begin{figure}
\centering
\includegraphics[width=0.85\textwidth]{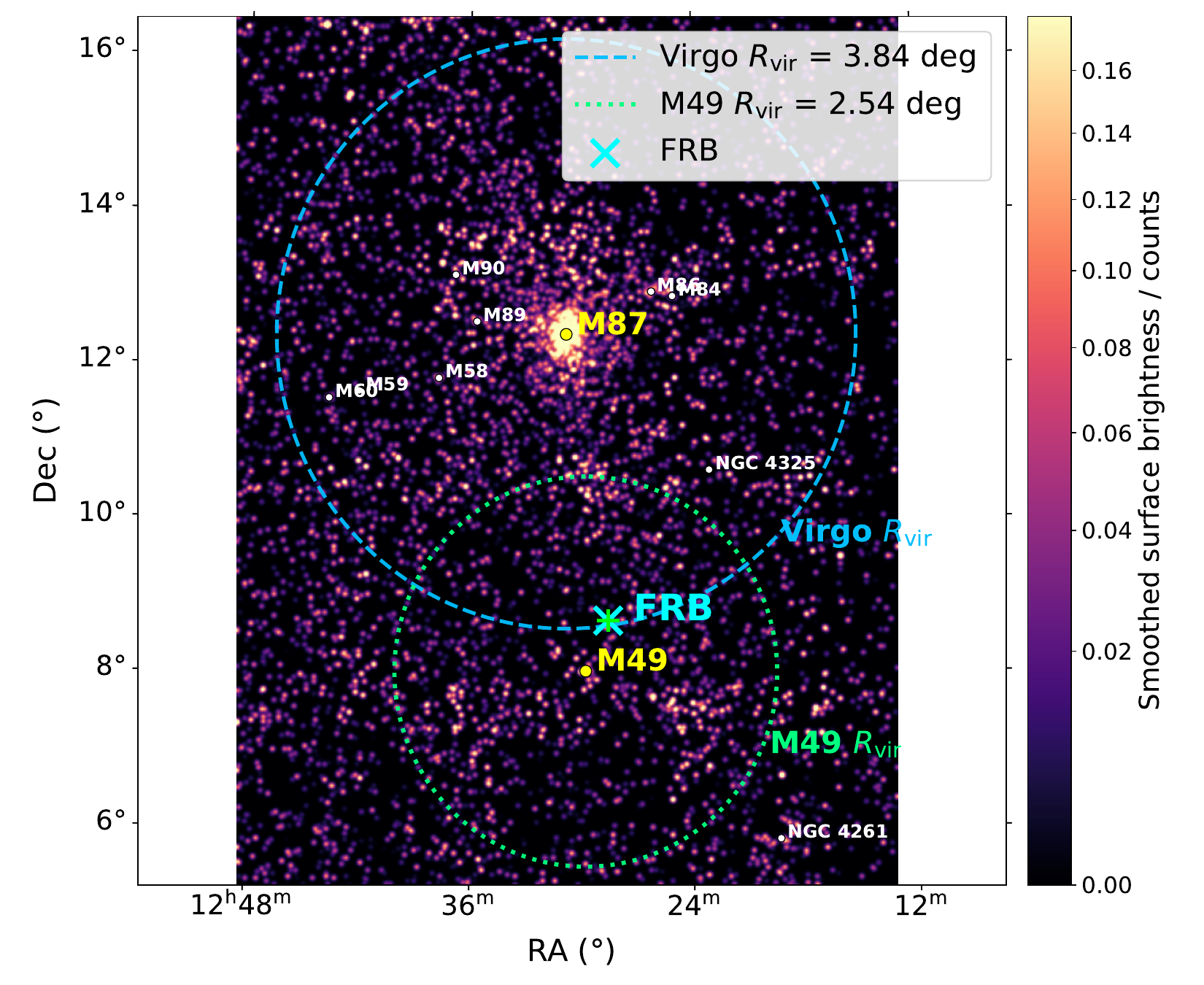}
\caption{
Foreground environment of FRB20230907D shown using a smoothed eROSITA X-ray surface brightness map \citep{Merloni2024,Bulbul2024,Kluge2024}.
The cyan dashed circle marks the adopted virial radius of the Virgo cluster while the green dotted circle marks the adopted virial radius of the M49 group.
The cyan cross marks the FRB position.
The sightline lies near the projected interface of the Virgo cluster and the M49 group.
}
\label{fig:frb20230907_erosita}
\end{figure}

Assuming that the host galaxy is at $z=0.464$, the FRB20230907D sightline intersects a complex low redshift foreground environment.
As illustrated by the eROSITA X-ray surface brightness map \citep{Merloni2024,Bulbul2024,Kluge2024} in Figure~\ref{fig:frb20230907_erosita}, the sightline lies in the projected overlap between the outskirts of the Virgo cluster and the M49 group.
For Virgo, we adopt $R_{\rm vir}\simeq r_{200}=1.08~{\rm Mpc}$, as inferred by \citet{Urban2011} from the mass--temperature scaling relation of \citet{Arnaud2005}; at the adopted distance of $16.1~{\rm Mpc}$, this corresponds to an angular radius of $3.84^\circ$.
For M49, we adopt the extrapolated virial radius $R_{\rm vir}=740~{\rm kpc}$ from \citet{Su2019}, corresponding to $2.54^\circ$ at its adopted distance of $16.7~{\rm Mpc}$.
With these radii, the FRB sightline lies in the projected interface between the Virgo and M49 environments.
Therefore, the burst may receive foreground DM contributions from several baryonic components, including diffuse gas in the Virgo/M49 region, intragroup or intracluster gas, and halo gas associated with nearby galaxies such as UGC~7596.

The foreground structures identified in this section define the components included in the subsequent DM budget analysis.
In Section~\ref{sec:dm_inference}, we combine the contributions from the diffuse IGM, the Virgo and M49 environments, the intervening foreground groups, and the FRB host, and infer the remaining contribution associated with UGC~7596.
The purpose of this decomposition is not to assign the observed DM uniquely to individual structures, but to assess how much of the large electron column along the FRB20230907D sightline can plausibly be accounted for by the identified foreground environment.

\subsection{Prime Focus Spectrograph (PFS) and Sloan Digital Sky Survey (SDSS) data}\label{sec:spectroscopic_data}

The principal data set used to anchor this analysis is a wide field spectroscopic survey obtained with the PFS on the Subaru Telescope \citep{Takada2014,Tamura2016}. 
The observations were conducted as part of the PFS Inference of FRB Lines of sight and Environments (PIFFLE) program under observing program ID S25A-039 (PI: Lee) on 2025 March 24 --- the second-ever night of science operations for PFS.
The field received approximately three hours of total integration distributed among several overlapping pointings, yielding spectra for 6464 individual targets.

\begin{figure}
\centering
\includegraphics[width=0.70\textwidth]{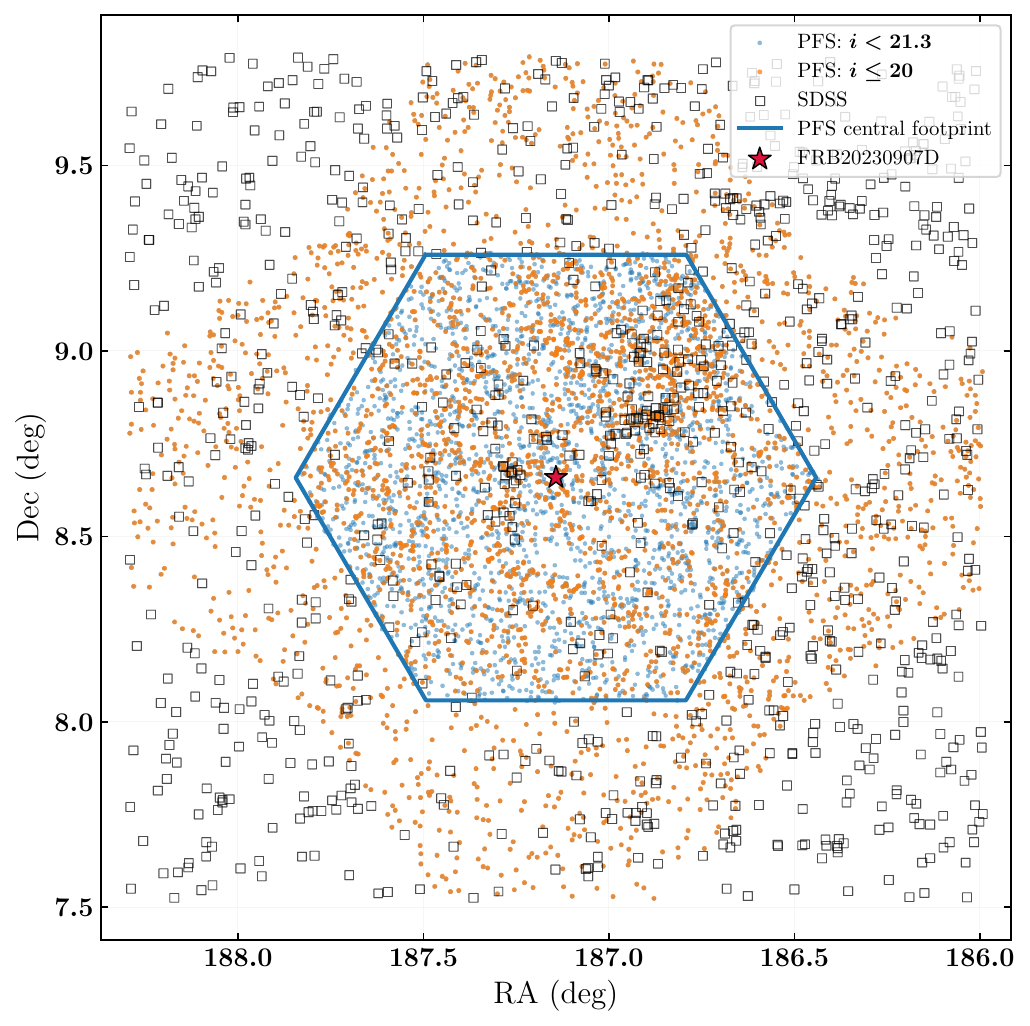}
\caption{
Angular distribution of the spectroscopic foreground sample used in the group analysis.
Blue points show targets with $i<21.3$, while orange points show the brighter subsample with $i\leq20$. 
Open squares show the supplementary SDSS spectroscopic sample.
The polygon marks the central PFS footprint, and the red star marks the position of FRB20230907D.
The concentration of bright PFS and SDSS galaxies near $({\rm RA},{\rm Dec})\simeq(187^\circ,8.8^\circ)$ is associated with the $z\simeq0.09$ foreground structure identified in our group analysis, which exhibits significant projected substructure (Section~\ref{sec:z009_system}).
}
\label{fig:pfs_target_position}
\end{figure}

The primary PFS targets were selected from the Pan-STARRS1 (PS1) photometric catalog \citep{Chambers2016,Flewelling2020}, primarily on the basis of apparent $r$-band magnitude.
Nominal cumulative exposure times of $900$, $1800$, and $2700~{\rm s}$ were assigned to targets satisfying $r_{\rm AB}<19.8$, $r_{\rm AB}<20.6$, and $r_{\rm AB}<21.3$, respectively. 
Within each predefined target category, the actual spectroscopic targets were selected randomly from the eligible galaxies.
Galaxies brighter than $r_{\rm AB}=17.5$ were excluded from the primary PFS target selection due to concerns about charge persistence to the near-infrared arm of the spectrograph, as well as the assumption that SDSS covers galaxies brighter than this. 
The central, approximately $1.25~{\rm deg}^{2}$, PFS footprint was covered by multiple pointings to the primary depth of $r_{\rm AB}<21.3$, while four shallower pointings targeting galaxies with $r_{\rm AB}<19.8$ were offset to the north, east, south, and west to extend the transverse coverage. 
The resulting pointing geometry and spatial sampling are shown in Figure~\ref{fig:pfs_target_position}.
The pointings overlapped within an approximately $50~{\rm arcmin}^{2}$ around the FRB position, where the galaxies could conceivably be directly intersected by the FRB sightline. Galaxies within this field were selected down to $i_{\rm AB}\lesssim23$, receiving cumulative exposure times of up to approximately $5400~{\rm s}$.
The resulting sampling therefore varies with both apparent magnitude and angular position and should not be regarded as volume limited.

The PFS observations were reduced by the Subaru/PFS Data Reduction Pipeline (DRP), which performs spectral extraction, wavelength calibration, sky subtraction, flux calibration, and coaddition of the individual exposures \citep{Tanaka2026}.
We use the observatory-provided reduced and calibrated one-dimensional spectra from the PFS \texttt{run21\_June2025} processing release, produced with \texttt{drp\_stella} version \texttt{PIPE2D-1703} (commit \texttt{7f169e3}), for the subsequent redshift measurements.
Spectroscopic redshifts were measured using the \texttt{Marz} redshifting software \citep{Hinton2016}. \footnote{We used the \texttt{Marz} implementation available at \url{https://github.com/Samreay/Marz}.}
The \texttt{Marz} solutions were also visually inspected.
A redshift quality flag, QOP, was assigned based on the confidence of the identification.
We define the secure redshift sample by requiring ${\rm QOP}\geq3$.
The PFS observing target selection and exposure tiers were defined using $r$-band magnitudes.
For the foreground and group finding analysis, however, we adopt a uniform $i$-band selection of $i_{\rm AB}<21.3$.
The redder $i$ band provides a more stable tracer of the galaxy population over the redshift range considered here, particularly at $z\gtrsim0.4$, where the $4000~\text{\AA}$ break enters the observed $r$ band.
Unless explicitly stated otherwise, magnitude cuts used in the subsequent foreground, group finding, and mock calibration analyses refer to $i$-band AB magnitudes.
After these selections, 5477 galaxies remain in the secure foreground sample.

To reproduce the PFS selection function used in Section~\ref{sec:validating_method}, we estimate the effective spectroscopic completeness relative to the same imaging catalog used to define the PFS target sample.
We select the imaging parent galaxy sample using the same angular selection and the magnitude limit $i_{\rm AB}<21.3$, and then positionally match the secure PFS redshift sample (${\rm QOP}\geq3$) to this parent sample.
The field averaged effective completeness is defined as
\begin{equation}
C_{\rm PFS}
=
\frac{N_{\rm imaging\rightarrow PFS}}
     {N_{\rm imaging,selected}},
\end{equation}
where $N_{\rm imaging,selected}$ is the number of imaging selected galaxies and $N_{\rm imaging\rightarrow PFS}$ is the number of these galaxies matched to a PFS spectrum satisfying ${\rm QOP}\geq3$. 
We obtain $C_{\rm PFS}=0.64$.
This quantity includes losses due to incomplete targeting, unsuccessful spectroscopy, and insecure redshift measurements, and is used when constructing the mock spectroscopic samples, used as reference for calibrating our group finding algorithm (see Section~\ref{sec:validating_method}).

Because the primary PFS selection excludes galaxies brighter than $r_{\rm AB}=17.5$, we supplement the PFS catalog with archival SDSS spectroscopy.
We first cross match the SDSS and PFS catalogs on the sky, treating sources separated by less than $2''$ as duplicate observations of the same galaxy and removing these duplicates from the SDSS sample.
The remaining 1007 SDSS galaxies are added to the secure PFS catalog to form the combined PFS+SDSS spectroscopic catalog, improving the sampling of bright foreground galaxies.
The addition of SDSS does not significantly alter the PFS selection completeness.
The supplementary SDSS galaxies are also shown in Figure~\ref{fig:pfs_target_position}, together with the PFS sample, to illustrate their spatial coverage across the field.

\begin{figure}
\centering
\includegraphics[width=0.98\textwidth]{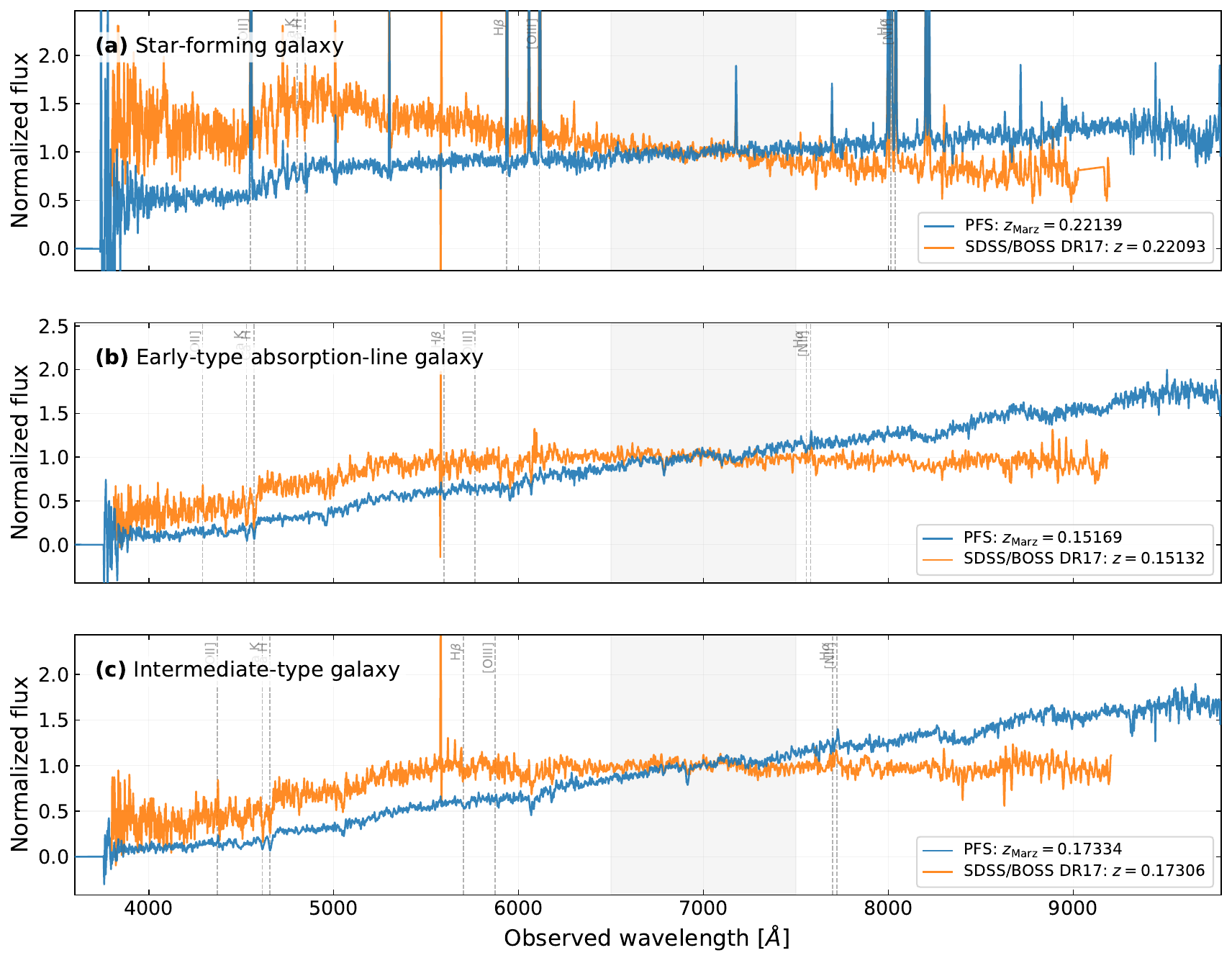}
\caption{
Comparison of continuum normalized PFS and SDSS/BOSS DR17 spectra for three randomly-selected galaxies in the FRB20230907D field. 
The blue and orange curves show the PFS and SDSS/BOSS spectra, respectively. 
The shaded regions indicate the wavelength intervals used for normalization, while the vertical dotted lines mark the expected observed wavelengths of prominent spectral features at the measured redshift. 
Although the continuum shapes and relative flux levels differ due to issues in the PFS spectrophotometric calibration, the spectral features occur at consistent wavelengths. 
}
\label{fig:pfs_spectra}
\end{figure}

To illustrate the consistency between the PFS and SDSS/BOSS DR17 redshift measurements, Figure~\ref{fig:pfs_spectra} compares the spectra of three galaxies observed by both surveys.
Although differences remain in the continuum shapes and relative flux calibration, the principal spectral features occur at consistent observed wavelengths. 
For these examples, the \texttt{Marz} and SDSS/BOSS redshifts agree to within $|\Delta z|\lesssim3\times10^{-4}$, which is sufficiently precise for the foreground group finding analysis.

\begin{figure}
\centering
\includegraphics[width=0.70\textwidth]{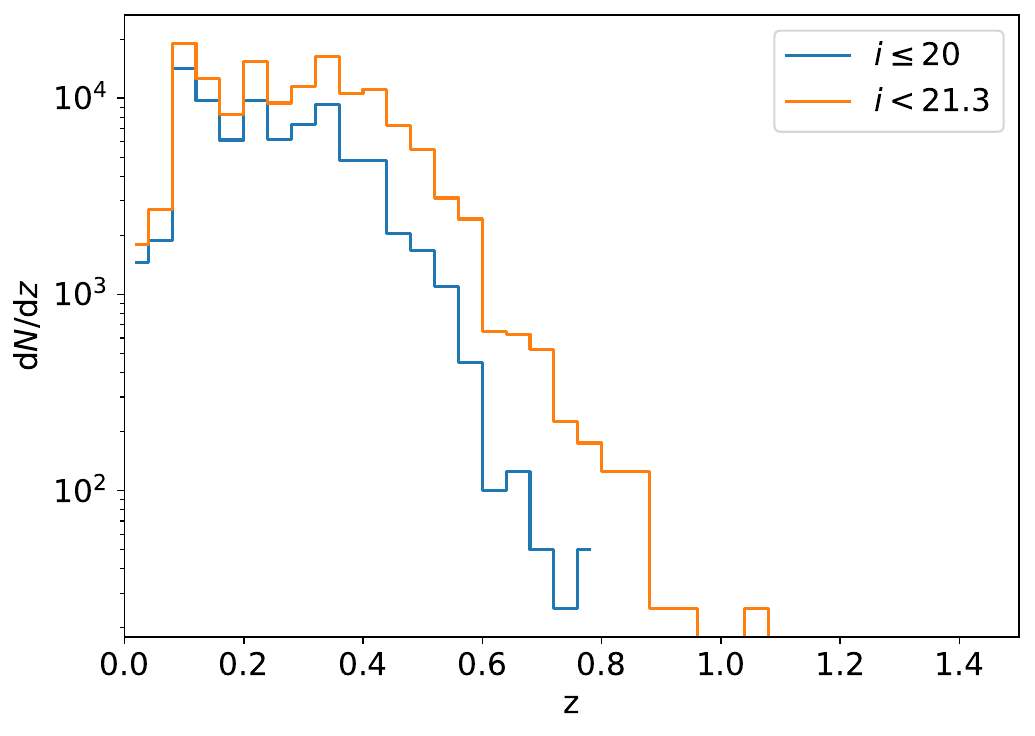}
\caption{
Redshift distribution of PFS galaxies for two $i$-band selections. 
The blue histogram shows galaxies with $i\leq20$, while the orange histogram shows galaxies with $i<21.3$. 
The $i<21.3$ sample provides denser sampling and extends to higher redshift. 
}
\label{fig:pfs_redshift_distribution}
\end{figure}

The redshift distribution of the secure PFS sample is shown in Figure~\ref{fig:pfs_redshift_distribution}.
The brighter $i_{\rm AB}\leq20$ subsample is preferentially concentrated at lower redshift, while the full $i_{\rm AB}<21.3$ sample provides denser coverage and extends to higher redshift.

\begin{figure}
\centering
\includegraphics[width=0.90\textwidth]{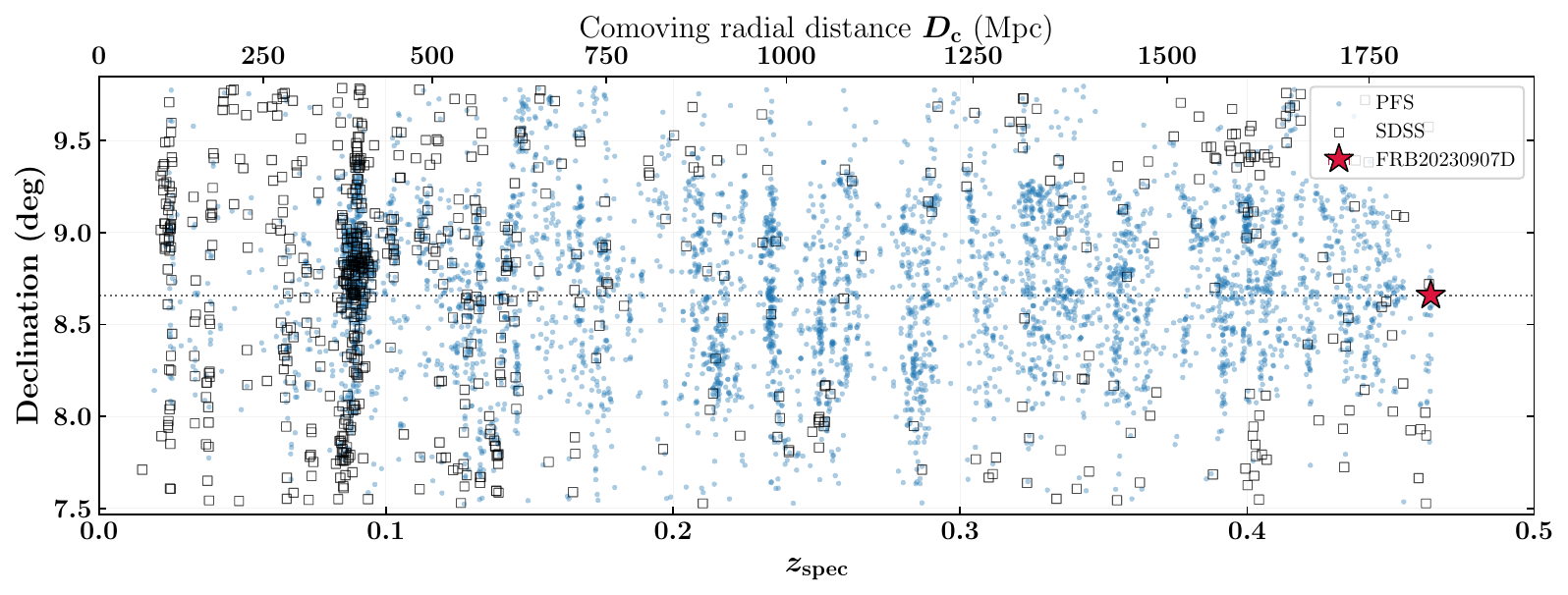}
\caption{
Declination and spectroscopic redshift distribution for foreground galaxies in the FRB20230907D field.
Blue points show PFS galaxies and open squares show the supplementary SDSS sample.
The red star marks position of FRB20230907D at $z_{\rm FRB}=0.464$ and the horizontal dotted line marks its declination.
}
\label{fig:z_declination}
\end{figure}

Figure~\ref{fig:z_declination} shows the declination and spectroscopic redshift distribution of foreground galaxies toward FRB20230907D.
Several narrow concentrations in redshift are visible, illustrating the large scale structures intersecting the FRB field.

\begin{figure}
\centering
\includegraphics[width=0.70\textwidth]
{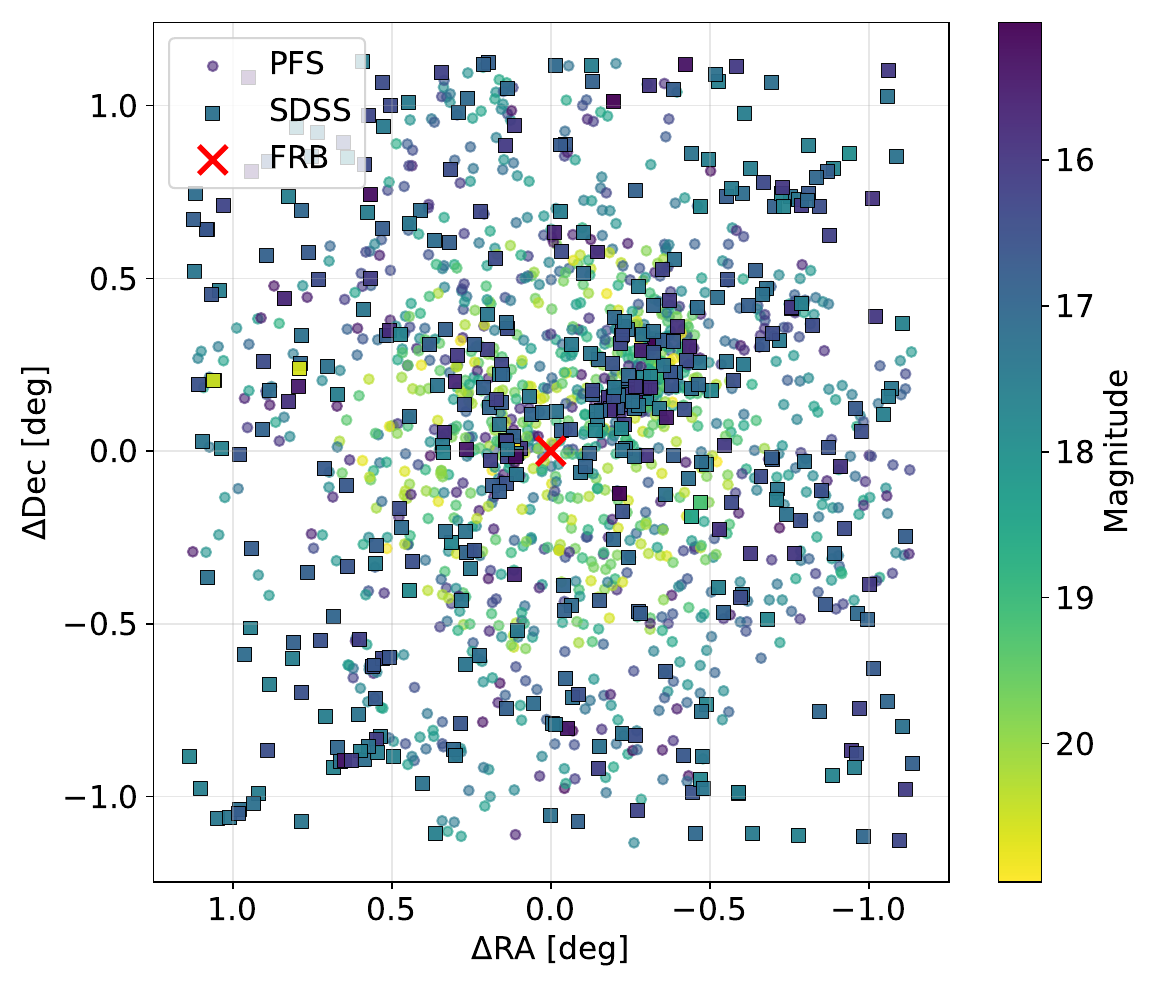}
\caption{
Projected distribution of the combined PFS and SDSS spectroscopic sample over $0.045<z<0.155$, shown relative to the FRB20230907D position. 
Circles show PFS galaxies, squares show the supplementary SDSS galaxies, and the red cross marks the FRB sightline.
The color scale indicates the $i$-band AB magnitude.
}
\label{fig:pfs_sdss_z01_positions}
\end{figure}

As seen in Figure~\ref{fig:z_declination}, the spectroscopic redshift distribution contains a prominent concentration of galaxies near $z\simeq0.09$.
To examine the projected environment of this structure, we select galaxies over $0.045<z<0.155$, corresponding to the lowest redshift interval adopted in our subsequent group finding analysis (Section~\ref{sec:validating_method}).
This interval contains 1316 PFS galaxies and 416 supplementary SDSS galaxies. 
Figure~\ref{fig:pfs_sdss_z01_positions} shows their projected distribution relative to the FRB sightline. 

For each galaxy in the combined catalog, the angular separation from the FRB determines its projected transverse position relative to the sightline, while the spectroscopic redshift determines its radial comoving distance.
These coordinates provide the input to the group finding procedure described in Section~\ref{sec:validating_method}.

\subsection{Friends-of-friends group finding}\label{sec:validating_method}

To identify galaxy groups and clusters in the foreground of FRB20230907D, we use a friends-of-friends (FoF) group finding approach.
The basic idea of a FoF algorithm is to link galaxies that are close to each other in projected position and redshift space.
Galaxies connected by the adopted linking lengths are assigned to the same group, and the resulting group catalog can then be used to estimate group redshifts, memberships, velocity dispersions, characteristic radii, and halo masses.

In this work, we use the Nessie\footnote{Nessie: \url{https://github.com/TrystanScottLambert/nessie_py}} group finder.
Nessie is a fast and flexible FoF galaxy group finder based on the GAMA group finding method of \citet{Robotham2011}.
It constructs galaxy group catalogs from redshift survey data using galaxy sky positions and redshifts as the main inputs.
The free parameters of the algorithm determine how aggressively galaxies are linked into groups.
In particular, the parameter $b_0$ controls the transverse linking length, while $r_0$ controls the line of sight linking length in redshift space.
A larger $b_0$ links galaxies over larger projected separations, while a larger $r_0$ allows galaxies with larger line of sight separations or velocity differences to be assigned to the same group.
Therefore, these parameters must be chosen carefully before applying the group finder to the FRB foreground field.

We calibrate these parameters using the mock lightcone catalogs of \citet{Henriques2015} (hereafter H15). 
The mock galaxy catalogs are extracted from the Millennium simulation and populated using a semi-analytic galaxy formation model in a Planck cosmology, providing both redshift space galaxy catalogs and the corresponding true halo memberships. 
We select the mock galaxies to reproduce the $i_{\rm AB}<21.3$ magnitude limit and approximate the effective PFS spectroscopic completeness measured in Section~\ref{sec:spectroscopic_data} by randomly downsampling the mock catalogs to $C_{\rm PFS}=0.64$.
We do not tune the mock catalogs to reproduce the detailed observed redshift distribution, which reflects the particular large scale structure along the sightline.
The calibration is performed independently in five slightly overlapping redshift bins centered at $z=0.1$, $0.2$, $0.3$, $0.4$, and $0.5$, with centers separated by $\Delta z=0.1$ and each bin having a width of $0.11$. 
The small $\Delta z=0.01$ overlap between adjacent bins is introduced to reduce boundary effects for systems lying close to the edges of the redshift intervals while maintaining a redshift-local calibration of the linking parameters.
In each bin, Nessie is run over a grid of $b_0$ and $r_0$.
The detailed mock--Nessie matching procedure, optimization criteria, and validation statistics are presented together with the calibration results in Section~\ref{sec:mock_validation_results}.

\section{Results}\label{sec:results}
\subsection{Mock validation results}
\label{sec:mock_validation_results}


For each trial parameter pair, the recovered Nessie groups are compared with the true mock halos. 
Candidate mock-Nessie pairs are first required to satisfy $|\Delta z|<0.02$ and $d_{\rm proj}<5~{\rm Mpc}$.
The pairs are then assigned using a greedy one-to-one matching procedure. 
The matching cost combines projected separation, redshift difference, and, when available, halo mass difference, thereby preventing one mock halo from being assigned to multiple recovered groups, or vice versa.

The optimized parameters are chosen by minimizing the total score
\begin{equation}
\label{equ:opt_score}
S_{\rm tot}
=
\lambda_{\rm pos}S_{\rm pos}
+
\lambda_z S_z
+
\lambda_M S_M
+
\lambda_N S_N
+
\lambda_{\rm unmatch}S_{\rm unmatch}
+
S_{\rm reg}.
\end{equation}
We adopt fixed weights $\lambda_{\rm pos}=1$, $\lambda_z=0.05$, $\lambda_M=1$, $\lambda_N=2$, and $\lambda_{\rm unmatch}=5$.
Here $S_{\rm pos}$, $S_z$, and $S_M$ quantify the differences in projected position, redshift, and halo mass for matched systems. 
The term $S_N$ penalizes differences between the numbers of true and recovered groups, while $S_{\rm unmatch}$ penalizes unmatched systems. 
The weak regularization term $S_{\rm reg}$ is included only to stabilize the optimization by preventing the internally estimated matching scales from becoming unrealistically small or large.
The redshift term is down-weighted so that small differences in group redshift do not dominate the total score.

After selecting the optimal parameters, listed in Table~\ref{tab:best_nessie_parameters_by_redshift}, we evaluate membership recovery using purity, completeness, and the F1 score as follows. 
For the $i$th matched mock Nessie group pair, let $N_{i}^{{\rm overlap}}$ be the number of galaxies that are members of both the true mock halo and its matched Nessie group, $N_i^{\rm Nessie,mem}$ the number of galaxies assigned to the recovered Nessie group, and $N_i^{\rm mock,mem}$ the number of true mock halo members. 
We define
\begin{equation}
P_i
=
\frac{N_i^{\rm overlap}}
     {N_i^{\rm Nessie,mem}},
\qquad
C_i
=
\frac{N_i^{\rm overlap}}
     {N_i^{\rm mock,mem}},
\end{equation}
and
\begin{equation}
F_{1,i}
=
\frac{2P_iC_i}{P_i+C_i}.
\end{equation}
Here $P_i$, $C_i$, and $F_{1,i}$ denote the purity, completeness, and F1 score, respectively \citep{Sokolova2009}.
Purity quantifies contamination in the recovered group, whereas completeness quantifies the recovered fraction of the true membership.
The F1 score is the harmonic mean of purity and completeness and is high only when both are high.
The values reported in Table~\ref{tab:nessie_mock_validation_mass_bins} are medians over matched mock halo--Nessie group pairs in bins of the true mock halo mass, with the bin boundaries listed in the table.

Nessie estimates a raw dynamical mass, $M_{\rm Nessie,raw}$, from the member galaxy velocity dispersion and projected distribution.
It then applies the luminosity function correction  \citep{Lambert2025}

\begin{equation}
M_{\rm Nessie}
=
F_{\rm LF}M_{\rm Nessie,raw},
\qquad
F_{\rm LF}
=
\frac{
\displaystyle\int_{-\infty}^{\infty}\Phi(M)\,{\rm d}M
}{
\displaystyle\int_{-\infty}^{M_{\rm lim}}\Phi(M)\,{\rm d}M
},
\end{equation}
where  $\Phi(M)$ is the galaxy luminosity function expressed as a function of absolute magnitude $M$, and $M_{\rm lim}$ is the absolute magnitude limit corresponding to the survey flux limit at the group redshift. 
The factor $F_{\rm LF}$ corrects for galaxies fainter than the survey limit.
We compare this corrected mass with the true mock halo mass through
\begin{equation}
\Delta\log M
=
\log_{10}M_{\rm Nessie}
-
\log_{10}M_{\rm mock}.
\end{equation}
We also calculate the three-dimensional positional offset, $\Delta d_{\rm 3D}$, between the recovered Nessie group center and the true mock halo center.

Table~\ref{tab:nessie_mock_validation_mass_bins} summarizes the recovery statistics in three halo mass bins. 
For matched systems, the median F1 scores range from $0.758$ to $0.842$, with median purity near $0.8$ in all three bins. 
The median mass offset is small for the two highest mass bins, although its scatter increases toward lower halo mass. 
The median positional offset is approximately $1.3~{\rm Mpc}$ for the most massive systems and decreases at lower mass.

\begin{table}
\centering
\caption{
Mock validation of the Nessie group finder in three halo mass bins.
The reported purity, completeness, F1 score, mass offset, and positional offset are medians over matched mock Nessie group pairs.
$N_{\rm mock,grp}$ and $N_{\rm Nessie,grp}$ denote the total numbers of independent groups in each mass bin.}
\label{tab:nessie_mock_validation_mass_bins}
\resizebox{\textwidth}{!}{
\begin{tabular}{lccccccccc}
\hline
Mass bin 
& $N_{\rm pair}$ 
& $F_1$ 
& $P$ 
& $C$ 
& $\Delta \log M$ 
& $\sigma_{\Delta \log M}$ 
& $d_{\rm 3D}$ 
& $N_{\rm mock,grp}$ 
& $N_{\rm Nessie,grp}$ \\
& 
& 
& 
& 
& 
& 
& [Mpc] 
& 
& \\
\hline
$\log_{10}(M/M_\odot)>14.5$ 
& 296 
& 0.758 
& 0.799 
& 0.802 
& 0.056 
& 0.326 
& 1.336 
& 307 
& 786 \\

$13.5 \leq \log_{10}(M/M_\odot)<14.5$ 
& 8436 
& 0.800 
& 0.818 
& 0.889 
& 0.098 
& 0.370 
& 0.996 
& 8747 
& 19921 \\

$12.5 \leq \log_{10}(M/M_\odot)<13.5$ 
& 17910 
& 0.842 
& 0.818 
& 1.000 
& 0.211 
& 0.523 
& 0.626 
& 20933 
& 24035 \\
\hline
\end{tabular}
}
\end{table}

Despite the good membership statistics for matched systems, Nessie recovers substantially more independent groups than are present in the mock catalog.
For $\log_{10}(M/M_\odot)>14.5$, it identifies 786 systems compared with 307 true groups, while in the intermediate mass bin it identifies 19921 systems compared with 8747 true groups. 
This does not conflict with the reported purity, completeness, and F1 scores, because those statistics are calculated only for successfully matched pairs and do not penalize unmatched Nessie detections.

The excess may arise from fragmentation of individual mock halos, spurious redshift space associations, and scatter in the recovered dynamical masses.
The present validation does not distinguish quantitatively among these effects. 
We therefore use Nessie to identify candidate foreground structures, rather than treating the recovered catalog as an unbiased one-to-one census of the halo population. 
Systems relevant to the FRB sightline are inspected individually in Section~\ref{sec:pfs_sdss_group_catalog}.

\begin{figure}
\centering
\includegraphics[width=0.99\textwidth]{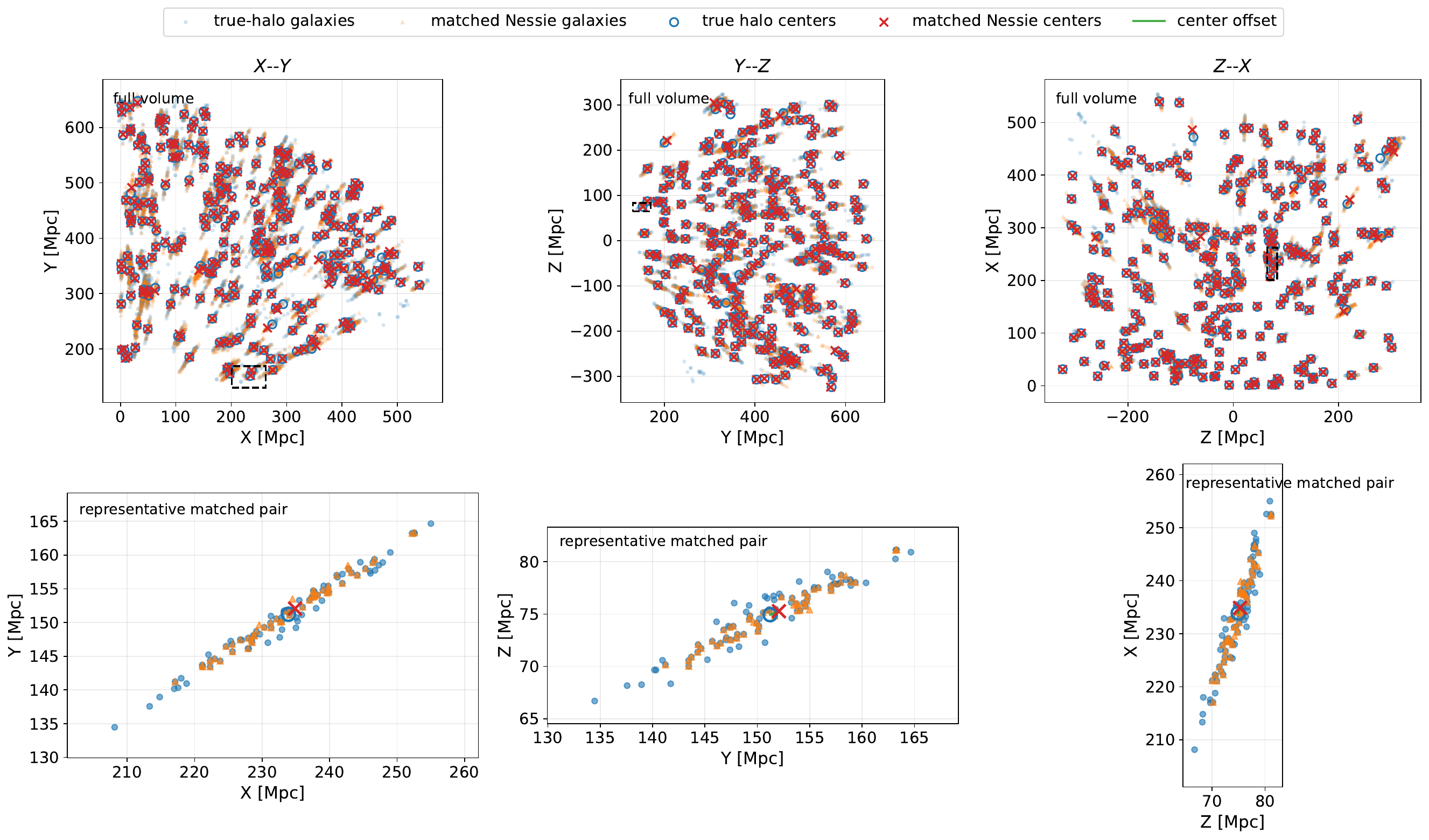}
\caption{
Mock validation of the Nessie group finder for massive systems with $\log_{10}(M/M_\odot)>14.5$.
The upper panels show the full $X$--$Y$, $Y$--$Z$, and $Z$--$X$ projections of the mock volume.
In the adopted coordinate system, $Y$ is the line-of-sight coordinate, while $X$ and $Z$ are the two transverse coordinates.
Blue circles show galaxies belonging to the true mock halos, while orange triangles show galaxies assigned to the matched Nessie groups.
Open blue circles mark the true mock halo centers, and red crosses mark the corresponding matched Nessie centers.
The dashed boxes indicate the representative system enlarged in the lower panels.
The lower panels show the same matched system in the three projections, with the green line indicating the offset between the true and recovered group centers.
The representative pair is selected to lie close to the median $F_1$ score and median $\Delta d_{\rm 3D}$ of the high-mass matched sample.
}
\label{fig:nessie_mock_validation_logM_gt_14p5}
\end{figure}

Figure~\ref{fig:nessie_mock_validation_logM_gt_14p5} compares the true and recovered systems in the highest mass bin, $\log_{10}(M/M_\odot)>14.5$.
The upper panels show the full mock volume in the $X$--$Y$, $Y$--$Z$, and $Z$--$X$ projections, demonstrating that the matched Nessie centers generally trace the same large scale concentrations as the true mock halo centers.
The lower panels provide an enlarged view of a representative matched system, selected to lie close to the median $F_1$ score and median $\Delta d_{\rm 3D}$ of the high mass sample.
This zoom illustrates both the close agreement of the recovered and true group centers and the residual differences in the assigned galaxy membership.

\subsection{Group catalog}
\label{sec:group_catalog}

\subsubsection{PFS+SDSS foreground catalog}
\label{sec:pfs_sdss_group_catalog}

We apply the mock calibrated Nessie group finder to the combined PFS+SDSS spectroscopic catalog in five overlapping redshift bins covering $0.045<z<0.555$.
For each bin, we use the optimized linking parameters derived from the mock calibration in Section~\ref{sec:mock_validation_results}, without further adjustment to the observed PFS+SDSS catalog.
Table~\ref{tab:best_nessie_parameters_by_redshift} lists the adopted parameters, the number of input galaxies, and the number of recovered groups with at least five members. 
Because the Nessie search begins at $z=0.045$, the nearby Virgo and M49 environments are not included and are considered separately in Section~\ref{sec:low_redshift_catalog}.

\begin{table}
\centering
\caption{
Best fit Nessie parameters used to construct the PFS+SDSS group catalog in each redshift bin. 
The values of $b_0$ and $r_0$ are the optimal parameters obtained from our H15 mock calibration by minimizing the score defined in Equation~\ref{equ:opt_score}.
The redshift bin boundaries are fixed by construction, $N_{\rm gal}$ gives the number of galaxies in each bin, and $N_{\rm group}$ gives the number of recovered groups with at least five members.
}
\label{tab:best_nessie_parameters_by_redshift}
\begin{tabular}{ccccccc}
\hline
$z_{\rm cen}$
& $z_{\rm min}$
& $z_{\rm max}$
& $b_0$
& $r_0$
& $N_{\rm gal}$
& $N_{\rm group}$ \\
\hline
0.100 & 0.045 & 0.155 & 0.08 & 15.0 & 1732 & 40 \\
0.200 & 0.145 & 0.255 & 0.06 & 20.0 & 1357 & 27 \\
0.300 & 0.245 & 0.355 & 0.04 & 25.0 & 1407 & 18 \\
0.400 & 0.345 & 0.455 & 0.04 & 20.0 & 1310 & 21 \\
0.500 & 0.445 & 0.555 & 0.04 & 20.0 & 783  & 6  \\
\hline
\end{tabular}
\end{table}

For each recovered group, we calculate its center, mean redshift, membership, projected separation from the FRB sightline, characteristic radius, and dynamical mass. 
We identify candidate foreground intersections by requiring the FRB impact parameter to satisfy $b<2R_{\rm vir}$, where $R_{\rm vir}$ is the virial radius inferred for each recovered group.
Within the PFS+SDSS search range, only one recovered system satisfies this criterion. 
It is the prominent foreground association near $z\simeq0.09$, which we examine in detail in Section~\ref{sec:z009_system}.

Table~\ref{tab:z0p1_massive_groups} lists the most massive groups recovered in the lowest Nessie redshift bin, $0.045<z<0.155$. 
The most massive initial association is Nessie group~1 at $z\simeq0.09$, with a mean redshift $z=0.0896$, 411 member galaxies, and an initial Nessie mass estimate $\log_{10}(M_{\rm Nessie}/M_\odot)=15.131$.
This mass is much larger than the masses of the other recovered systems in the same redshift bin. 
Although Nessie group 1 is clearly the most important candidate foreground halo in the PFS+SDSS redshift range, its initial membership, radius, and mass require further inspection before being used in the DM analysis (see Section~\ref{sec:intervening_group_dm}).

\begin{table}
\centering
\caption{
Most massive Nessie groups in the $z_{\rm cen}=0.1$ redshift bin. 
The table lists the group rank by inferred mass, Nessie group ID, group center, redshift, member counts, characteristic radius, and inferred halo mass.
}
\label{tab:z0p1_massive_groups}
\resizebox{\textwidth}{!}{
\begin{tabular}{cccccccccc}
\hline
Rank
& Group ID
& RA
& Dec
& $z$
& $N_{\rm tot}$
& $N_{\rm PFS}$
& $N_{\rm SDSS}$
& $R_{\rm Nessie}$
& $\log_{10}(M_{\rm Nessie}/M_\odot)$ \\
&
& [deg]
& [deg]
&
&
&
&
& [Mpc]
& \\
\hline
1  & 1   & 186.9666 & 8.8396 & 0.08955 & 411 & 325 & 86 & 4.476 & 15.131 \\
2  & 39  & 187.5016 & 8.5700 & 0.12860 & 5   & 5   & 0  & 1.297 & 14.405 \\
3  & 140 & 187.2825 & 9.5832 & 0.14721 & 13  & 12  & 1  & 2.125 & 14.348 \\
4  & 148 & 186.5683 & 9.2431 & 0.09048 & 7   & 6   & 1  & 1.030 & 14.279 \\
5  & 180 & 186.3775 & 9.0831 & 0.10052 & 15  & 11  & 4  & 1.160 & 14.274 \\
6  & 94  & 188.0302 & 8.7958 & 0.12483 & 6   & 5   & 1  & 0.893 & 14.230 \\
7  & 173 & 186.2736 & 8.5309 & 0.14258 & 8   & 7   & 1  & 0.782 & 14.060 \\
8  & 6   & 187.6129 & 8.9795 & 0.12153 & 8   & 7   & 1  & 1.070 & 14.050 \\
9  & 123 & 186.7226 & 8.8955 & 0.11391 & 6   & 6   & 0  & 1.587 & 14.004 \\
10 & 81  & 187.1830 & 9.1379 & 0.11308 & 6   & 6   & 0  & 0.723 & 13.898 \\
\hline
\end{tabular}
}
\end{table}

\subsubsection{The \texorpdfstring{$z\simeq0.09$}{z~0.09} foreground system}
\label{sec:z009_system}

The $z\simeq0.09$ foreground association identified by Nessie has counterparts in previously published group catalogs based on SDSS spectroscopy. 
In the SDSS DR12 based catalog of \citet{Saulder2016}, the corresponding system is group 65627, centered at $({\rm RA},{\rm Dec})=(186.8750^\circ,8.8249^\circ)$, with redshift $z=0.090192$. 
The catalog assigns 110 galaxies to the system and reports a line of sight velocity dispersion $\sigma_{v,{\rm Saulder}}=743.5~{\rm km~s^{-1}}$.
\citet{Saulder2016} calculated the velocity dispersion from the catalog membership using a gapper based estimator, including a finite sample correction and an adopted SDSS redshift uncertainty of $30~{\rm km~s^{-1}}$ \citep{Beers1990,Eke2004,Saulder2016}. 
The catalog also reports a characteristic radius $R_{\rm cat}=1.056~{\rm Mpc}$ and a calibrated total mass $\log_{10}(M_{\rm tot}/M_\odot)=15.1626$. 

An independent counterpart is present in the SDSS DR10 based catalog of \citet{Tempel2014}. 
The corresponding system is group 7689, centered at $({\rm RA},{\rm Dec})=(186.96421^\circ,8.82234^\circ)$, with CMB frame redshift $z_{\rm cmb}=0.09048$. 
The catalog assigns 107 galaxies to the system and reports $\sigma_{v,{\rm Tempel}}=747.7~{\rm km~s^{-1}}$.

The close agreement in position and redshift indicates that the Nessie, \citet{Saulder2016}, and \citet{Tempel2014} detections correspond to the same physical foreground structure. 
However, the initial Nessie association contains 411 galaxies, has a characteristic radius of $4.476~{\rm Mpc}$, and has a inferred mass $\log_{10}(M_{\rm Nessie}/M_\odot)=15.131$. 
Its larger number of members relative to the SDSS-based catalogs is expected from the denser PFS spectroscopic sampling.
However, the unusually large characteristic radius and group mass motivate further inspection of the initial association.
\citet{PastorMarazuela2025} previously noted three foreground cluster candidates from the DESI Legacy Imaging Surveys catalog of \citet{Zou2021}.
Two are associated with the $z\simeq0.09$ foreground complex.
ID~2074100062 has $z_{\rm spec}=0.089421$ and lies close to the retained main concentration, whereas ID~2074100051 has $z_{\rm spec}=0.086638$ and lies in the spatially offset component removed above.
The FRB sightline passes ID~2074100051 at $b\simeq0.88~{\rm Mpc}$, just outside its catalog $R_{500}=0.815~{\rm Mpc}$.

\begin{figure}
\centering
\includegraphics[width=0.95\textwidth]
{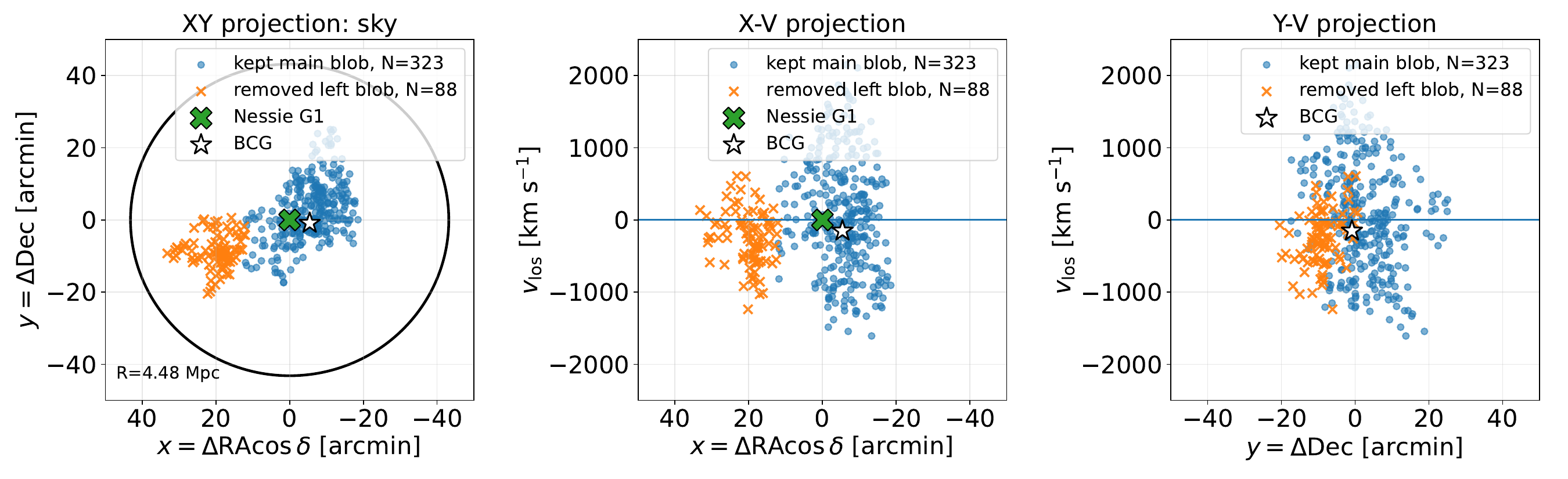}
\caption{
Visualizations of the most massive Nessie group in the $z\simeq0.09$ foreground structure. 
The left panel shows the sky distribution relative to the group center, while the middle and right panels show projected position--velocity diagrams. 
The blue points show the main component retained for the final group estimate, while the orange points show the removed projected component. 
The green cross marks the original Nessie group center, and the star marks the visually selected candidate BCG.
}
\label{fig:nessie_group1_blob_removal}
\end{figure}

Figure~\ref{fig:nessie_group1_blob_removal} shows the spatial and velocity distribution of the initially assigned Nessie members. 
Visual inspection of the projected distribution reveals two distinct concentrations: a dominant component near the recovered group center and a smaller, spatially offset component. 
We separate the latter using a cut at $x_{\rm proj}=12'$ relative to the Nessie group center, chosen to lie in the projected gap between the two concentrations.
The cleaned system contains 323 galaxies, including 260 PFS and 63 SDSS members. 
We compute the mean spectroscopic redshift of the retained galaxies, obtaining $z_{\rm group}=0.089806$.

\begin{figure}
\centering
\includegraphics[width=0.70\textwidth]
{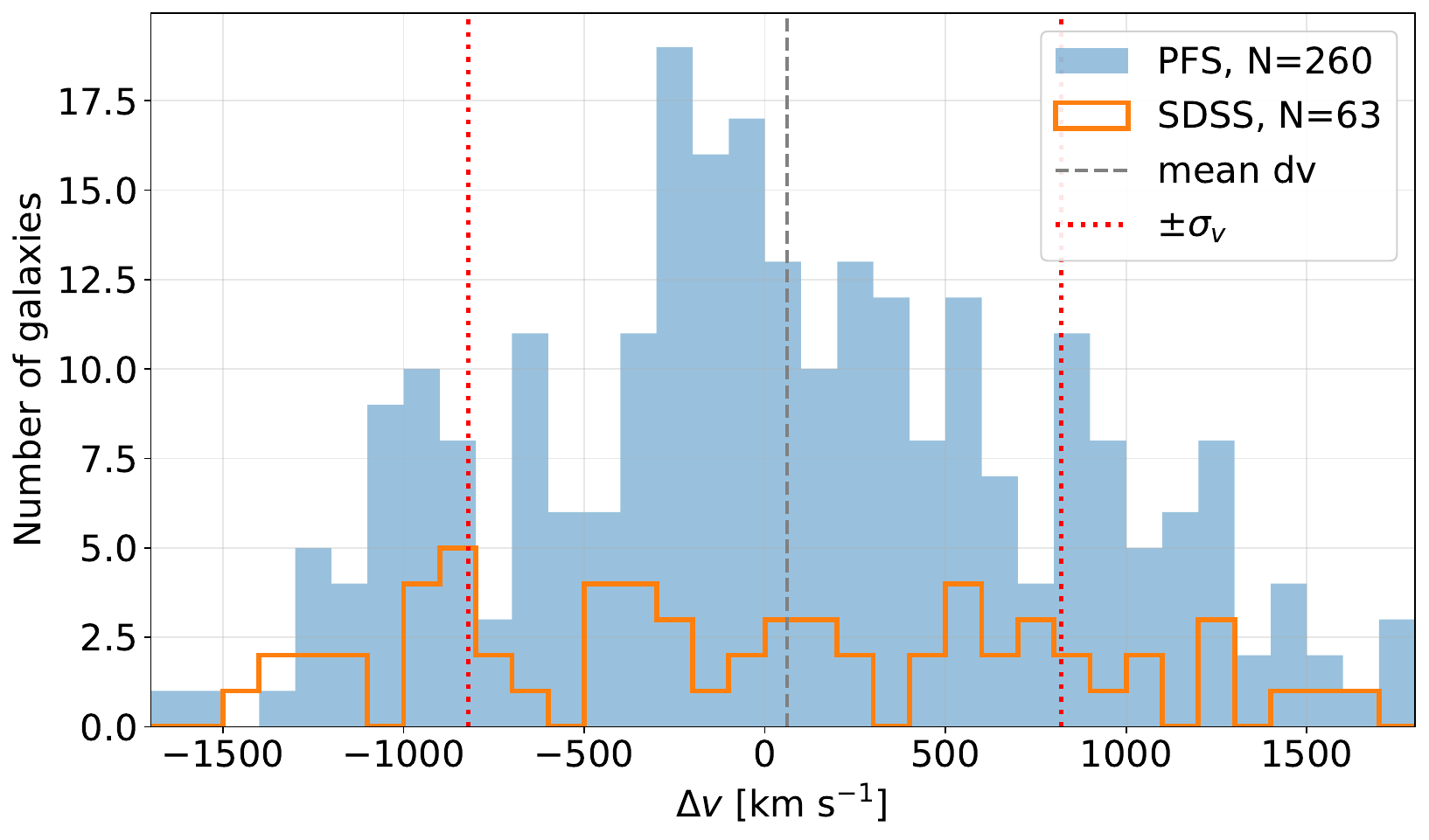}
\caption{
Line of sight velocity distribution of the cleaned $z\simeq0.09$ group after removing the secondary projected component. 
The blue histogram shows the PFS galaxies, while the orange histogram shows the SDSS galaxies. 
The vertical markers indicate the mean velocity and the corresponding $\pm\sigma_v$ interval.
}
\label{fig:velocity_after_blob_removal}
\end{figure}

Figure~\ref{fig:velocity_after_blob_removal} compares the line of sight velocity distributions of the cleaned PFS and SDSS subsamples. 
The SDSS subsample is more sparsely sampled, but both subsamples trace a similar global velocity distribution.
The cleaned value, $\sigma_v\simeq776~{\rm km~s^{-1}}$, agrees to within a few percent with the \citet{Saulder2016} and \citet{Tempel2014} measurements.
The denser PFS sampling primarily provides a more detailed view of the projected and velocity structure, allowing the group membership to be refined.

As an independent dynamical mass estimate, we use the velocity dispersion--mass relation of \citet{Evrard2008},
\begin{equation}
\sigma_{\rm DM}(M,z)
=
\sigma_{15}
\left[
\frac{h(z)M_{200}}{10^{15}~M_\odot}
\right]^\alpha,
\end{equation}
where $\sigma_{15}=1082.9~{\rm km~s^{-1}}$, $\alpha=0.3361$, and $h(z)=H(z)/(100~{\rm km~s^{-1}~Mpc^{-1}})$.
Treating the cleaned one-dimensional galaxy velocity dispersion as a proxy for the dark matter velocity dispersion gives
\begin{equation}
M_{200,\sigma}
=
\frac{10^{15}~M_\odot}{h(z)}
\left(
\frac{\sigma_v}{\sigma_{15}}
\right)^{1/\alpha}
\simeq
5.2\times10^{14}~M_\odot.
\end{equation}

The use of member galaxy velocities as a proxy for the dark matter dispersion may introduce systematic uncertainty from velocity bias, residual interlopers, and departures from dynamical equilibrium.
Nevertheless, the cleaned velocity dispersion is consistent with the independent measurements of \citet{Saulder2016} and \citet{Tempel2014}, despite their substantially smaller spectroscopic memberships.
The PFS+SDSS sampling allows us to identify and remove projected substructure before estimating the velocity dispersion and halo mass.
The resulting $M_{200,\sigma}$ is substantially below the initial Nessie mass estimate and also below the catalog mass reported by \citet{Saulder2016}; the latter difference additionally reflects differences in membership selection, mass definition, and mass calibration.
For our fiducial mass estimate, we instead apply the simulation-based inference (SBI) framework of \citet{Wang2026}, retrained on Uchuu mock catalogs matched to the redshift and observational selection of the PFS sample (Appendix~\ref{app:sbi}).
The resulting posterior has a mean $\langle M_{\rm vir}\rangle=7.7\times10^{14}~M_\odot$, with a 16--84 percentile range of $(6.38$--$8.99)\times10^{14}~M_\odot$.

We also search for counterparts of the $z\simeq0.09$ foreground structure in independent optical and X-ray data.
In the DESI Legacy Imaging Surveys imaging \citep{Dey2019}, we identify a prominent extended galaxy near the projected center of the adopted group center as a candidate brightest cluster galaxy (BCG).
Its position and spectroscopic redshift are
\begin{equation}
({\rm RA},{\rm Dec})_{\rm BCG}
=
(186.843967^\circ,8.840906^\circ),
\qquad
z_{\rm BCG}=0.090228.
\end{equation}
Because the object was selected visually and no quantitative luminosity ranking was performed, we use it only as a tracer of the galaxy concentration and not as the adopted halo center.

\begin{figure}
\centering
\includegraphics[width=\textwidth]
{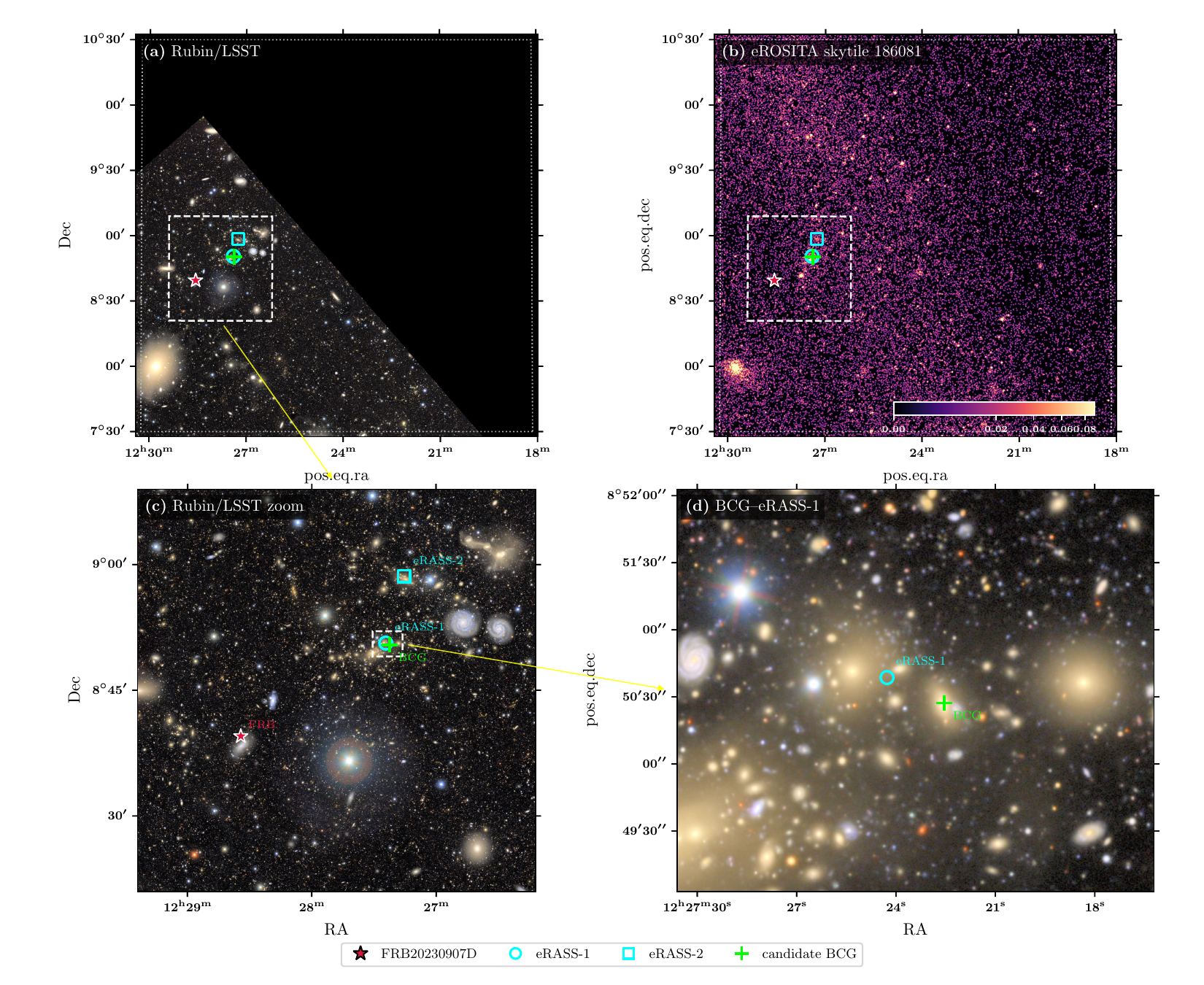}
\caption{
Multiscale optical and X-ray view of the $z\simeq0.09$ foreground structure.
Panel (a) shows the wide-field Rubin/LSST image, with the positions of FRB20230907D, eRASS-G1, eRASS-G2, and the candidate BCG marked.
Panel (b) shows the full eROSITA skytile 186081 X-ray image over the same broad region.
The white dashed rectangles in panels (a) and (b) indicate the region shown in the intermediate Rubin/LSST zoom in panel (c), while the white dashed rectangle in panel (c) indicates the region enlarged in panel (d).
Panel (d) shows a tighter view of the candidate BCG and eRASS-G1.
The spatial association between the optical galaxy overdensity and the X-ray emission supports the presence of a massive foreground hot gas halo.
}
\label{fig:rubin_erosita_environment}
\end{figure}

Figure~\ref{fig:rubin_erosita_environment} provides a multiscale optical and X-ray view of the $z\simeq0.09$ foreground structure \citep{Merloni2024,Bulbul2024,Kluge2024}.
The wide field Rubin/LSST and eROSITA panels show the relative positions of FRB20230907D, the candidate BCG, and two eROSITA sources, eRASS~J122724.2+085038 (hereafter eRASS-G1) and eRASS~J122715.3+085836 (hereafter eRASS-G2). 
The lower panels provide progressively closer optical views of the foreground system, with the tightest view highlighting the projected association between the candidate BCG and eRASS-G1. 
The spatial correspondence between the optical galaxy overdensity and the X-ray emission independently supports the presence of a massive foreground hot gas halo.

The catalog properties of the two X-ray detections are
\begin{equation}
({\rm RA},{\rm Dec})_{\rm eRASS-G1}
=
(186.85113^\circ,8.84407^\circ),
\qquad
z_{\rm eRASS-G1}\simeq0.0897,
\qquad
M_{500,{\rm eRASS-G1}}\simeq3.1\times10^{14}~M_\odot,
\end{equation}
and
\begin{equation}
({\rm RA},{\rm Dec})_{\rm eRASS-G2}
=
(186.81415^\circ,8.97694^\circ),
\qquad
z_{\rm eRASS-G2}\simeq0.0893,
\qquad
M_{500,{\rm eRASS-G2}}\simeq7.3\times10^{13}~M_\odot.
\end{equation}

The eROSITA masses are reported as $M_{500}$, whereas the dynamical estimate above is expressed as $M_{200}$.
For an NFW halo \citep{Navarro1997} with a representative cluster scale concentration, the approximate conversion $M_{200}\simeq1.4M_{500}$ gives $M_{200,{\rm eRASS-G1}}\simeq4.3\times10^{14}~M_\odot$, for the primary X-ray source, broadly consistent with the velocity dispersion estimate. 
This conversion depends on halo concentration and cosmology and is used only as an approximate comparison \citep{Hu2003}.

The offsets among the X-ray centroids, optical group center, and candidate BCG suggest that the system may be dynamically complex or not fully relaxed. 
We therefore use the X-ray detections as independent evidence for a massive hot gas halo, but do not adopt their centroids or masses as the fiducial parameters in the DM calculation.

For the subsequent DM calculation, we adopt the cleaned Nessie center, mean redshift, and the SBI virial mass posterior:
\begin{equation}
({\rm RA},{\rm Dec})_{\rm halo}
=
(186.9666^\circ,8.8396^\circ),
\qquad
z_{\rm halo}=0.089806,
\qquad
M_{\rm vir, halo}
=
7.70\times10^{14}~M_\odot.
\end{equation}
The FRB sightline lies $\theta_{\rm FRB}\simeq0.251^\circ$ from the adopted group center, corresponding to a projected impact parameter of $b\simeq1.57~{\rm Mpc}$.
For the mean SBI mass, the adopted halo model gives $R_{\rm vir}=2.3~{\rm Mpc}$ and $b/R_{\rm vir}\simeq0.68$.

Table~\ref{tab:z009_catalog_comparison} summarizes the independent optical, spectroscopic, and X-ray measurements associated with the $z\simeq0.09$ foreground system. 
The Saulder, Tempel, Zou, candidate BCG, and eROSITA measurements provide independent confirmation of the foreground system, but we will define the fiducial halo parameters to be used in our analysis in  Section~\ref{sec:intervening_group_dm}.

\begin{table*}
\centering
\caption{
Summary of measurements associated with the $z\simeq0.09$ foreground system. 
Masses retain the definitions reported by the corresponding catalog or analysis; quantities such as $M_{\rm Nessie}$, $M_{\rm tot}$, $M_{200}$, $M_{\rm vir}$, and $M_{500}$ are therefore not directly equivalent and are shown primarily for cross identification and consistency checks.
}
\label{tab:z009_catalog_comparison}

\resizebox{\textwidth}{!}{
\begin{tabular}{lccccccc}
\hline
Tracer or catalog
& RA
& Dec
& $z$
& $N_{\rm mem}$
& $\sigma_v$
& Radius
& Mass \\
& [deg]
& [deg]
&
&
& [${\rm km~s^{-1}}$]
& 
& \\
\hline
Nessie, initial
& 186.9666
& 8.8396
& 0.08955
& 411
& \nodata
& $R_{\rm Nessie}=4.476~{\rm Mpc}$
& $\log_{10}(M_{\rm Nessie}/M_\odot)=15.131$ \\

PFS+SDSS, cleaned
& 186.9666
& 8.8396
& 0.089806
& 323
& 776
& $R_{\rm vir}=2.3~{\rm Mpc}$
& $M_{\rm vir}=7.70\times10^{14}~M_\odot$ \\

\citet{Saulder2016}
& 186.8750
& 8.8249
& 0.090192
& 110
& 743.5
& $R_{\rm cat}=1.056~{\rm Mpc}$
& $\log_{10}(M_{\rm tot}/M_\odot)=15.1626$ \\

\citet{Tempel2014}
& 186.96421
& 8.82234
& 0.09048
& 107
& 747.7
& $R_{\rm vir}=0.680~h^{-1}{\rm Mpc}$
& $M_{\rm NFW}=7.71\times10^{14}\,h^{-1}M_\odot$ \\

\citet{Zou2021}, ID 2074100062
& 186.91014
& 8.82134
& 0.089421
& \nodata
& \nodata
& $R_{500}=1.026~{\rm Mpc}$
& $\log_{10}(M_{500}/M_\odot)=14.514$ \\

\citet{Zou2021}, ID 2074100051
& 187.28598
& 8.69091
& 0.086638
& \nodata
& \nodata
& $R_{500}=0.815~{\rm Mpc}$
& $\log_{10}(M_{500}/M_\odot)=14.217$ \\

Candidate BCG
& 186.843967
& 8.840906
& 0.090228
& \nodata
& \nodata
& \nodata
& \nodata \\

eRASS-G1
& 186.85113
& 8.84407
& 0.0897
& \nodata
& \nodata
& \nodata
& $M_{500}=3.1\times10^{14}~M_\odot$ \\

eRASS-G2
& 186.81415
& 8.97694
& 0.0893
& \nodata
& \nodata
& \nodata
& $M_{500}=7.3\times10^{13}~M_\odot$ \\
\hline
\end{tabular}
}

\end{table*}

\subsubsection{Low redshift foreground catalog}
\label{sec:low_redshift_catalog}

Because the PFS+SDSS Nessie search is restricted to $z>0.045$, it might not probe the lower redshift foreground that might lie at relatively larger ($\gtrsim 1\,\deg$ angular separations from the FRB sightline. 
We therefore inspect the published group catalogs of \citet{Saulder2016}, \citet{Tempel2014}, and \citet{Tully2015} for systems whose projected group regions approach the FRB sightline. 
The search recovers the nearby Virgo region associations, which are treated separately through the Virgo and M49 gas models in Section~\ref{sec:dm_inference}. 
It also identifies a distinct foreground group at $z\simeq0.02565$ (hereafter the $z\simeq0.026$ group), well separated in redshift from both the local Virgo/M49 environment and the $z\simeq0.09$ system.

\citet{PastorMarazuela2025} also noted a lower redshift cluster candidate from the \citet{Zou2021} catalog, ID~2125300112.
This source has $z_{\rm phot}=0.010049$ has no catalog spectroscopic redshift, and is substantially lower than the spectroscopic redshift of the $z\simeq0.026$ group identified here.
Given that $z\sim 0.01$ lies well within the Local Volume that is thoroughly covered by spectroscopic observations, we assume this candidate is spurious and do not consider it any further.

In the \citet{Saulder2016} catalog, this system has group ID 36336, with $({\rm RA},{\rm Dec})=(186.617905^\circ,9.019892^\circ)$ and $z=0.025649$.
The catalog assigns 52 galaxies to the group and reports a velocity dispersion of $\sigma_v=364.1~{\rm km~s^{-1}}$ and a characteristic
group radius of $R_{\rm Saulder}=923.4~{\rm kpc}$, corresponding to an angular
radius $\theta_{\rm Saulder}=0.5170^\circ$.
Its calibrated total mass estimate is $M_{\rm tot}=1.52\times10^{14}~M_\odot$.
The angular separation between this group center and the FRB sightline is $\theta_{\rm FRB}=0.6320^\circ$, corresponding to $\frac{\theta_{\rm FRB}}{\theta_{\rm Saulder}}
\simeq1.22$.
The sightline therefore lies outside the reported catalog radius but within twice that radius.

Counterparts of the same redshift space overdensity are also present in the \citet{Saulder2016} and \citet{Tully2015} catalogs. 
Because these catalogs use different galaxy samples, group finding methods, distance estimates, and characteristic radius definitions, their reported properties are not expected to agree exactly.
In particular, the Tully catalog reports the projected second turnaround radius, $R_{2t}$, rather than $R_{200}$ or a Bryan--Norman virial radius.
The catalog radii and masses should therefore be treated as definition dependent measurements rather than directly comparable estimates of a single spherical overdensity halo.

For the subsequent analysis, we adopt from \citet{Saulder2016} this group center, redshift, and calibrated total mass estimate as the fiducial catalog properties of this foreground system $M_{\rm tot} =1.52\times10^{14}~M_\odot$.
The correspondence between this catalog mass and the spherical overdensity mass required by the halo gas model is discussed in Section~\ref{sec:intervening_group_dm}.

\section{DM inference}
\label{sec:dm_inference}

Having identified the foreground structures along the FRB20230907D sightline, we now estimate their contributions to the total ${\rm DM}_{\rm obs}=1031~{\rm pc~cm^{-3}}$ \citep{PastorMarazuela2025}.
For this sightline, we write the DM budget as
\begin{equation}
\begin{split}
{\rm DM}_{\rm obs}
={}&
{\rm DM}_{\rm MW}
+
{\rm DM}_{\rm UGC7596}
+
{\rm DM}_{\rm Virgo}
+
{\rm DM}_{\rm M49}
\\
&+
{\rm DM}_{z\simeq0.026}
+
{\rm DM}_{z\simeq0.09}
+
{\rm DM}_{\rm IGM}
+
\frac{{\rm DM}_{\rm host}}{1+z_{\rm FRB}}.
\end{split}
\label{eq:dm_budget}
\end{equation}

Here ${\rm DM}_{\rm Virgo}$ and ${\rm DM}_{\rm M49}$ denote the contributions from the nearby Virgo cluster and M49 group environments.
The terms ${\rm DM}_{z\simeq0.026}$ and ${\rm DM}_{z\simeq0.09}$ correspond to the two intervening foreground systems identified in Section~\ref{sec:group_catalog}. 
The contribution associated with UGC~7596 is inferred below from the residual of the full budget, rather than being fixed by an independent halo model.
Because the individual components have different observational and modeling uncertainties, we infer the DM budget probabilistically. 
The resulting values should be interpreted as model dependent component estimates rather than unique measurements of the electron column associated with each structure.

\subsection{Intervening foreground group contributions}
\label{sec:intervening_group_dm}

We include the cleaned $z\simeq0.09$ system and the independently identified $z\simeq0.026$ group as the two fiducial intervening group components in the DM budget.

\begin{figure}
\centering
\includegraphics[width=0.98\textwidth]{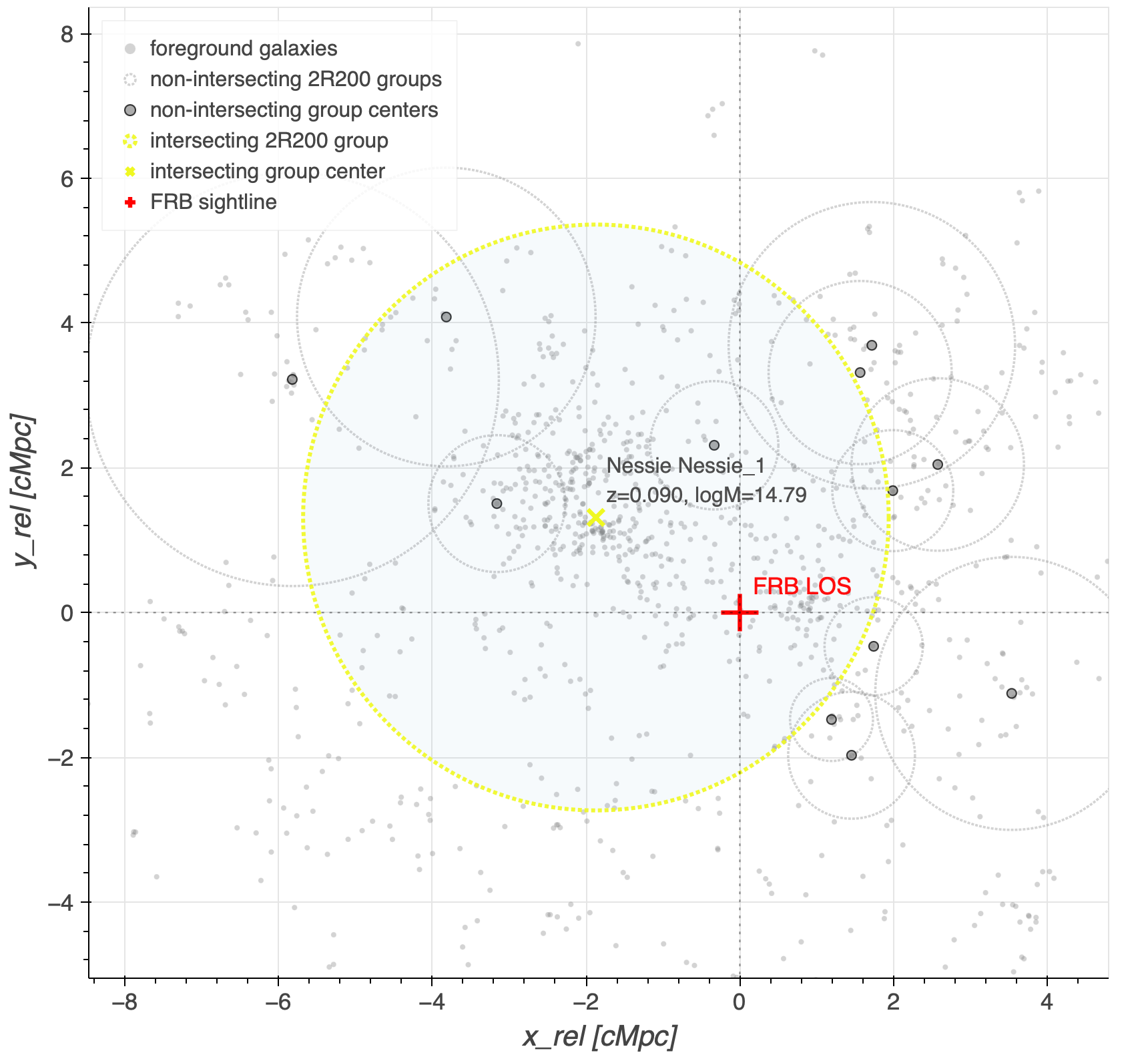}
\caption{
Foreground Nessie group intersections in the PFS+SDSS search range, $0.045<z<0.555$.
Grey points show foreground galaxies, grey circles show non-intersecting Nessie group regions, and the yellow circle marks the $z\simeq0.09$ system whose adopted $2R_{\rm vir}$ region intersects the FRB sightline.
The red cross marks the FRB line of sight.
}
\label{fig:intersect_halos_dm}
\end{figure}

Figure~\ref{fig:intersect_halos_dm} shows the projected Nessie group catalog and highlights the $z\simeq0.09$ system intersecting the sightline. 
The Virgo, M49, and $z\simeq0.02565$ systems are not shown because they lie outside the Nessie redshift selection.

We model the ionized gas associated with the intervening groups using the modified-NFW profile adopted by \citet{Prochaska2019},
\begin{equation}
\rho_{\rm b}(r)
=
\frac{\rho_{0,\rm b} f_{\rm hot}}
{y^{1-\alpha}(y_0+y)^{2+\alpha}},
\qquad
y=\frac{cr}{R_{\rm vir}}.
\end{equation}
Here $R_{\rm vir}$ is the halo virial radius, $c$ is the halo concentration, and $\alpha$ and $y_0$ determine the shape of the modified-NFW profile.
$f_{\rm hot}$ specifies the fraction of the halo's expected baryon mass,$(\Omega_{\rm b}/\Omega_{\rm m})M_{\rm halo}$, that resides in the diffuse, ionized, approximately virialized intragroup or intracluster gas.
The normalization $\rho_{0,\rm b}$ is set such that, before applying the factor $f_{\rm hot}$, the profile contains the cosmic baryonic budget
\begin{equation}
M_{\rm b}
=
\frac{\Omega_{\rm b}}{\Omega_{\rm m}}M_{\rm halo}.
\end{equation}
Thus, the gas mass represented by the modified-NFW component is $f_{\rm hot}M_{\rm b}$.
Assuming a fully ionized primordial H/He composition, we convert the gas mass density to the free electron number density as
\begin{equation}
n_e(r)
=
1.1667\,
\frac{\rho_{\rm b}(r)}
{1.33\,m_p},
\end{equation}
where $m_p$ is the proton mass.

For a halo at redshift $z_{\rm h}$ intersected by the FRB sightline at projected impact parameter $b$, its contribution to the observed DM is
\begin{equation}
{\rm DM}_{\rm halo}(b)
=
\frac{1}{1+z_{\rm h}}
\int_{-l_{\rm max}}^{+l_{\rm max}}
n_e\!\left(\sqrt{b^2+l^2}\right)\,{\rm d}l,
\qquad
l_{\rm max}
=
\sqrt{(2R_{\rm vir})^2-b^2}.
\end{equation}
The gas profile is truncated at $2R_{\rm vir}$, and the contribution is set to zero for $b>2R_{\rm vir}$.
For both intervening groups, we adopt $f_{\rm hot}=0.75$, $c=7.67$, $\alpha=2$, and $y_0=2$ following \citet{Prochaska2019,Khrykin2024a}.

To propagate the uncertainty in halo mass, we use $N_{\rm MC}=5\times10^{5}$ Monte Carlo realizations. 
For the $z\simeq0.09$ system, we draw the halo mass directly from the SBI posterior described in Appendix~\ref{app:sbi}.
For the $z\simeq0.02565$ group, we draw $\log_{10}M_{\rm halo}$ from a Gaussian distribution centered on the adopted fiducial halo mass with a standard deviation of $0.3$ dex.
For every realization, we recompute $R_{\rm vir}$ and the density normalization $\rho_{0,\rm b}$ for the sampled halo mass, while holding $b$, $f_{\rm hot}$, $c$, $\alpha$, and $y_0$ fixed, and repeat the line-of-sight DM calculation.

Applying this framework to the two intervening foreground groups, we obtain:
\begin{table*}
\centering
\caption{
Adopted properties and inferred DM contributions of the two intervening foreground groups. 
}
\label{tab:intervening_group_dm}
\begin{tabular}{lcccccccc}
\hline
System
& $z_{\rm h}$
& RA
& Dec
& $M_{\rm halo,fid}$
& $b$
& $R_{\rm vir}$
& $b/R_{\rm vir}$
& ${\rm DM}_{\rm halo}$ \\
&
& [deg]
& [deg]
& [$M_\odot$]
& [Mpc]
& [Mpc]
&
& [$\mathrm{pc\,cm^{-3}}$] \\
\hline
$z\simeq0.02565$
& 0.025649
& 186.617905
& 9.019892
& $1.52\times10^{14}$
& 1.215
& 1.385
& 0.88
& $80^{+60}_{-40}$ \\

$z\simeq0.09$
& 0.089806
& 186.9666
& 8.8396
& $7.70\times10^{14}$
& 1.566
& 2.3
& 0.68
& $205^{+30}_{-25}$ \\
\hline
\end{tabular}
\end{table*}

\begin{itemize}

\item \textit{The $z\simeq0.02565$ foreground group.}
For this system, we adopt the calibrated mass in \citet{Saulder2016}, $M_{\rm halo,fid}=1.52\times10^{14}~M_\odot$, as a proxy for the halo mass.
The FRB sightline passes the group at $b/R_{\rm vir}\simeq0.88$, from which we estimate ${\rm DM}_{z\simeq0.026}=80^{+60}_{-40}~{\rm pc~cm^{-3}}$.
Because the Saulder mass is a calibrated catalog total mass rather than an explicitly defined spherical overdensity mass, this estimate is subject to an additional mass definition systematic not fully represented by the quoted statistical uncertainty interval.

\item \textit{The $z\simeq0.09$ foreground group.}
For the cleaned PFS+SDSS system, we adopt the SBI virial mass posterior, whose mean is $M_{\rm vir}=7.70\times10^{14}~M_\odot$, with a 16--84 percentile range of $(6.38$--$8.99)\times10^{14}~M_\odot$.
For the mean mass, the sightline passes the halo at $b/R_{\rm vir}\simeq0.68$.
Propagating the full SBI mass posterior gives ${\rm DM}_{z\simeq0.09}=205^{+30}_{-25}~{\rm pc~cm^{-3}}$.
Additional systematic uncertainties arise from the adopted hot gas fraction, concentration, profile shape, spherical geometry, and truncation radius.
While $f_{\rm hot}$ mainly acts as an overall normalization of the predicted DM, the other assumptions can affect the inferred column density in a more sightline dependent manner.
\end{itemize}

\subsection{Other foreground and host contributions}
\label{sec:adopted_dm_terms}
The remaining foreground and host contributions to the DM budget are summarized below, proceeding from the local environment toward the FRB:

\begin{itemize}

\item \textit{Milky Way.}
For the MW contribution, we separate the disk/ISM and halo terms.
For the FRB20230907D sightline, we adopt ${\rm DM}_{\rm MW,ISM}=29~{\rm pc~cm^{-3}}$ from the NE2001 model \citep{PastorMarazuela2025}, and ${\rm DM}_{\rm MW,halo}=50\pm20~{\rm pc~cm^{-3}}$ for the Galactic halo \citep{Prochaska2019}.
We hold the ISM term fixed and draw the halo contribution from a Gaussian distribution in each Monte Carlo realization.
The resulting total MW contribution is ${\rm DM}_{\rm MW}=80\pm20~{\rm pc~cm^{-3}}$.

\item \textit{Virgo environment.}
For the Virgo environment, we adopt the power law electron density profile \citep{Planck2016}
\begin{equation}
n_e(r)
=
8.5\times10^{-5}
\left(\frac{r}{\rm Mpc}\right)^{-\beta}
{\rm cm^{-3}},
\qquad
\beta=1.21\pm0.12.
\end{equation}
We integrate the profile at $b_{\rm Virgo}=1.17~{\rm Mpc}$ over $-2.4<l<2.4~{\rm Mpc}$. 
Propagating the uncertainty in $\beta$ gives ${\rm DM}_{\rm Virgo}=226^{+13}_{-12}~{\rm pc~cm^{-3}}$.
This estimate is model dependent because the Virgo environment is spatially structured, whereas the calculation assumes a smooth spherical profile.

\item \textit{M49 group.}
We model M49 separately from the main Virgo component. 
The FRB sightline passes the group at $b_{\rm M49}\simeq0.211~{\rm Mpc}$. 
We adopt
\begin{equation}
M_{\rm halo,M49}=4.6\times10^{13}~M_\odot,
\qquad
f_{\rm g,M49}=\frac{M_{\rm gas}}{M_{\rm halo}}=0.006,
\end{equation}
following the halo mass estimate of \citet{Su2019} and the enclosed gas and total masses discussed by \citet{Schindler1999}.
Here $f_{\rm g,M49}$ is an absolute gas-to-total mass fraction, not a fraction of the cosmological baryon allotment.
We use the same modified-NFW profile shape and the same truncation radius, $2R_{\rm vir}$, as for the intervening groups.
The observed X-ray extent of approximately $300~{\rm kpc}$ provides evidence for stripping and truncation but is not used as the numerical integration limit. 
Assigning $\sigma_{\log_{10}M_{\rm M49}}=0.3~{\rm dex}$ and recalculating the contribution for each mass realization gives ${\rm DM}_{\rm M49}=11^{+4}_{-3}~{\rm pc~cm^{-3}}$.
This result is specific to the adopted absolute gas fraction and modified-NFW profile.

\item \textit{Diffuse IGM.}
For the diffuse IGM, we adopt $f_{\rm IGM}=0.64^{+0.09}_{-0.07}$ \citep{Khrykin2024b}, where gas assigned to identified foreground halos is modeled separately.
The mean contribution is calculated as \citep{Deng2014}
\begin{equation}
\left\langle{\rm DM}_{\rm IGM}(z)\right\rangle
=
\frac{3cH_0\Omega_{\rm b}f_{\rm IGM}}
     {8\pi Gm_{\rm p}}
\int_0^z
\frac{\chi_e(1+z')}
     {E(z')}
\,{\rm d}z',
\end{equation}
where $E(z)=\left[\Omega_{\rm m}(1+z)^3+\Omega_\Lambda\right]^{1/2}$ and $\chi_e=0.875$, appropriate for fully ionized hydrogen and doubly ionized helium.
At the FRB redshift, this gives ${\rm DM}_{\rm IGM}=310^{+40}_{-30}~{\rm pc~cm^{-3}}$.
We represent this uncertainty with a split normal distribution. 
It propagates only the uncertainty in $f_{\rm IGM}$ and does not include an additional sightline-to-sightline cosmic variance term.

\item \textit{Host galaxy.}
For the host contribution, we use the empirical relations between ${\rm DM}_{\rm host}$ and host galaxy stellar mass and star formation rate (SFR) reported by \citet{Bernales2025}.
Their fits give
\begin{equation}
{\rm DM}_{\rm host}^{(M_\star)} = 86 + 43\log_{10}
\left(
\frac{M_\star}{10^{10}M_\odot}
\right)
~{\rm pc~cm^{-3}},
\end{equation}
and
\begin{equation}
{\rm DM}_{\rm host}^{({\rm SFR})}
=
84
+
36\log_{10}
\left(
\frac{{\rm SFR}}{M_\odot\,{\rm yr}^{-1}}
\right)
~{\rm pc~cm^{-3}}.
\end{equation}
For FRB20230907D, we adopt $\log_{10}(M_\star/M_\odot)=10.9\pm0.2$ and ${\rm SFR}=14.9~M_\odot\,{\rm yr^{-1}}$, with a $0.3$ dex uncertainty in the SFR \citep{PastorMarazuela2025}.
The central relations give rest frame estimates of ${\rm DM}_{\rm host}^{(M_\star)}=125~{\rm pc~cm^{-3}}$ and ${\rm DM}_{\rm host}^{({\rm SFR})}=126~{\rm pc~cm^{-3}}$.
In each Monte Carlo realization, we draw the host properties and the coefficients of the two empirical relations from their quoted uncertainties.
Because the stellar mass and SFR estimates are not independent host DM measurements, we average the two predicted contributions.
We additionally include a $30\%$ multiplicative systematic uncertainty following \citet{Bernales2025}.
The rest frame contribution is converted to the observer frame using
\begin{equation}
{\rm DM}_{\rm host,obs}
=
\frac{{\rm DM}_{\rm host}}{1+z_{\rm FRB}}.
\end{equation}
The resulting observer-frame distribution has ${\rm DM}_{\rm host,obs} =80^{+30}_{-20}~{\rm pc~cm^{-3}}$.
\end{itemize}

Table~\ref{tab:dm_budget_inputs} summarizes the contributions and uncertainty distributions adopted for all components entering the Monte Carlo DM budget. 
The detailed halo geometries and gas model assumptions for the two intervening groups are given in Section~\ref{sec:intervening_group_dm}.

\begin{table*}
\centering
\caption{
Components entering the Monte Carlo DM budget for FRB20230907D.
Quoted values are medians with 16--84 percentile intervals unless otherwise stated. 
}
\label{tab:dm_budget_inputs}
\begin{tabular}{lccc}
\hline
Component
& Adopted input or model
& Uncertainty treatment
& ${\rm DM}$
\\
&
&
&
[${\rm pc~cm^{-3}}$]
\\
\hline
MW
& NE2001 ISM + MW halo
& ISM fixed; Gaussian MW-halo uncertainty
& $80\pm20$
\\

Diffuse IGM
& $f_{\rm IGM}=0.64^{+0.09}_{-0.07}$
& Split normal in $f_{\rm IGM}$
& $310^{+40}_{-30}$
\\

Virgo
& Power law electron density profile
& Gaussian uncertainty in $\beta$
& $226^{+13}_{-12}$
\\

M49
& Modified-NFW; $f_{\rm g}=0.006$
& $0.3$ dex halo mass scatter
& $11^{+4}_{-3}$
\\

$z\simeq0.02565$ group
& Modified-NFW; Saulder mass proxy
& $0.3$ dex halo mass scatter
& $80^{+60}_{-40}$
\\

$z\simeq0.09$ group
& Modified-NFW; SBI virial mass posterior
& SBI posterior samples
& $205^{+30}_{-25}$
\\

Host, observed frame
& Stellar mass and SFR scaling relations
& Monte Carlo + $30\%$ systematic
& $80^{+30}_{-20}$
\\
\hline
\end{tabular}
\end{table*}

\subsection{UGC 7596 DM contribution}
We constrain the DM contribution associated with UGC~7596 in two complementary ways.
First, we infer it as the residual of the observed DM after subtracting the MW, diffuse IGM, Virgo/M49, two intervening clustering, and host contributions, propagating their uncertainties through a Monte Carlo calculation.
We then compare this residual inference with independent constraints on the gas associated with UGC~7596, including the optically traced disk and an illustrative CGM halo model.

\subsubsection{Residual DM inference for UGC~7596}
\label{sec:ugc_residual}

Adopting the distance $D\simeq7.78~{\rm Mpc}$ reported by \citet{Daz-Garca2016}, the FRB sightline passes UGC~7596 at a projected separation of $b\simeq2.64~{\rm kpc}$.
Given that the galaxy has a stellar mass $M_\star\simeq1.57\times10^{8}~M_\odot$ \citep{Daz-Garca2016} and adopting the dwarf galaxy stellar-to-halo mass relation of \citet{Brook2014}, this corresponds to an indicative halo mass of $M_{\rm halo}\simeq3.5\times10^{10}~M_\odot$.
Using these galaxy and halo properties, \citet{PastorMarazuela2025} previously estimated a DM contribution of order $50~{\rm pc~cm^{-3}}$ from UGC~7596.
Ionized gas associated with its disk, ISM, or CGM could therefore contribute to the observed DM. 
Because the gas content and electron density profile of this low mass galaxy are poorly constrained, we first study how much DM the global budget allows to be associated with UGC~7596, without assuming a specific gas model for UGC~7596 itself.

The residual is defined as
\begin{equation}
\begin{split}
{\rm DM}_{\rm UGC7596}
={}&
{\rm DM}_{\rm obs}
-
{\rm DM}_{\rm MW}
-
{\rm DM}_{\rm IGM}
-
{\rm DM}_{\rm Virgo}
-
{\rm DM}_{\rm M49}
\\
&-
{\rm DM}_{z\simeq0.026}
-
{\rm DM}_{z\simeq0.09}
-
\frac{{\rm DM}_{\rm host}}{1+z_{\rm FRB}}.
\end{split}
\label{eq:ugc_residual}
\end{equation}
This residual inherits the uncertainties and model assumptions of every subtracted component and may also absorb unmodeled foreground gas. 
It should therefore not be interpreted as a direct measurement of the electron column physically associated with UGC~7596.

Using $N_{\rm MC}=5\times10^{5}$ Monte Carlo realizations, we draw the uncertain component contributions from the distributions summarized in  Table~\ref{tab:dm_budget_inputs} and evaluate Equation~\ref{eq:ugc_residual}  realization by realization. 
For the MW, the ISM contribution is held fixed while the halo contribution  is varied according to its adopted uncertainty. 
The uncertainties in the diffuse IGM, Virgo environment, M49 group, two intervening groups, and host contribution are likewise propagated through their adopted distributions.

\begin{figure}
\centering
\includegraphics[width=0.72\textwidth]{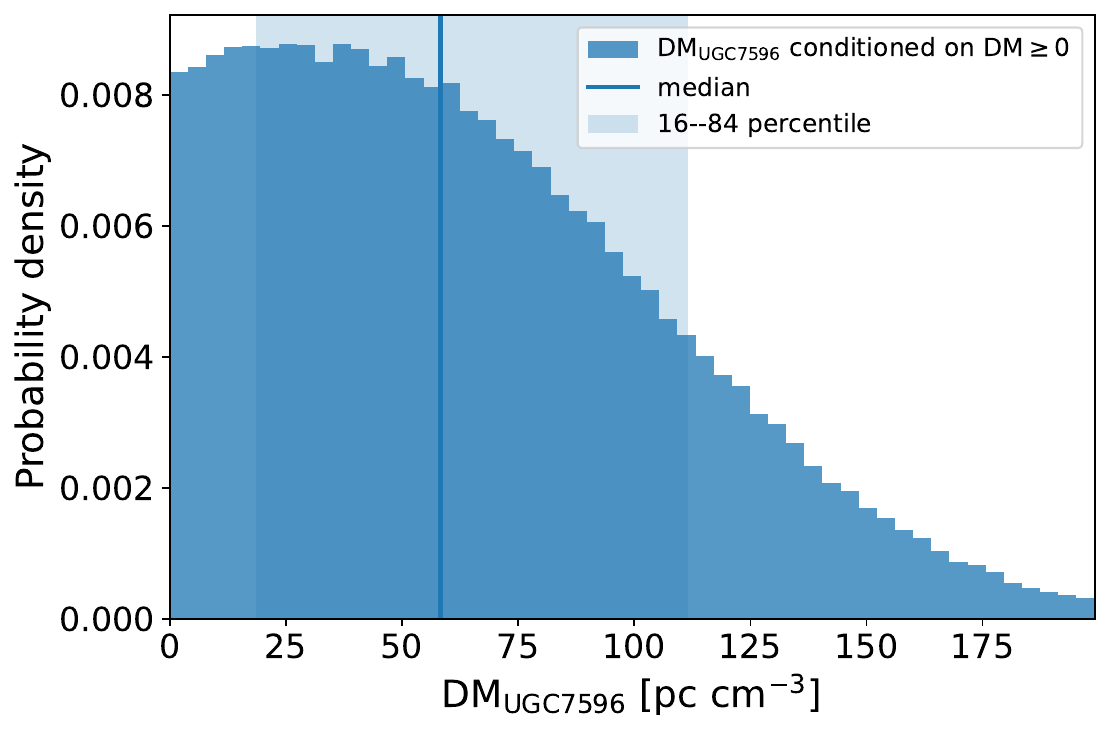}
\caption{
Residual DM associated with UGC~7596, conditional on a non-negative value.
The distribution is obtained by subtracting Monte Carlo realizations of the other terms in the FRB20230907D DM budget. 
The solid line marks the conditional median and the shaded region encloses the conditional $16$--$84$ percentile interval.
The conditional median is ${\rm DM}_{\rm UGC7596}=58~{\rm pc~cm^{-3}}$, and $57.9\%$ of the unconditioned realizations have a non-negative residual.
}
\label{fig:ugc_dm_inference}
\end{figure}

Before imposing a non-negative physical boundary, the residual distribution is ${\rm DM}_{\rm UGC7596,raw} = 16^{+73}_{-86}~{\rm pc~cm^{-3}}$, where the quoted uncertainties enclose the $16$--$84$ percentile interval.
The interval crosses zero, and $57.9\%$ of the Monte Carlo realizations yield a non-negative residual.
Because a physical electron column cannot be negative, we also examine the distribution conditional on ${\rm DM}_{\rm UGC7596,raw}\geq0$.
The resulting conditional distribution is ${\rm DM}_{\rm UGC7596} = 58^{+53}_{-40}~{\rm pc~cm^{-3}}$.
This conditional distribution characterizes the allowed magnitude of the residual when it is non-negative, but it should not be interpreted as evidence that a positive UGC~7596 contribution is required.
Figure~\ref{fig:ugc_dm_inference} shows the residual distribution conditional on a non-negative value. 
The truncation at zero and extended upper tail reflect the broad uncertainties in the foreground and host terms.

The raw residual is consistent with zero, so the present DM budget does not require a distinct positive contribution from UGC~7596. 
Only slightly more than half ($57.9\%$) of the Monte Carlo realizations yield a non-negative residual.
The conditional distribution permits a contribution of several tens of ${\rm pc~cm^{-3}}$, with its upper 68-percentile range extending to approximately $110~{\rm pc~cm^{-3}}$.
Such a residual could arise partly from the ISM or the CGM of UGC~7596, but it could also reflect uncertainties in the diffuse IGM, Virgo and M49 models, the two intervening groups, the host contribution, or unmodeled foreground gas. 
We therefore treat the residual as a model dependent budget remainder rather than a direct detection of gas associated with UGC~7596.

\begin{figure*}
    \centering
    \includegraphics[width=0.95\linewidth]{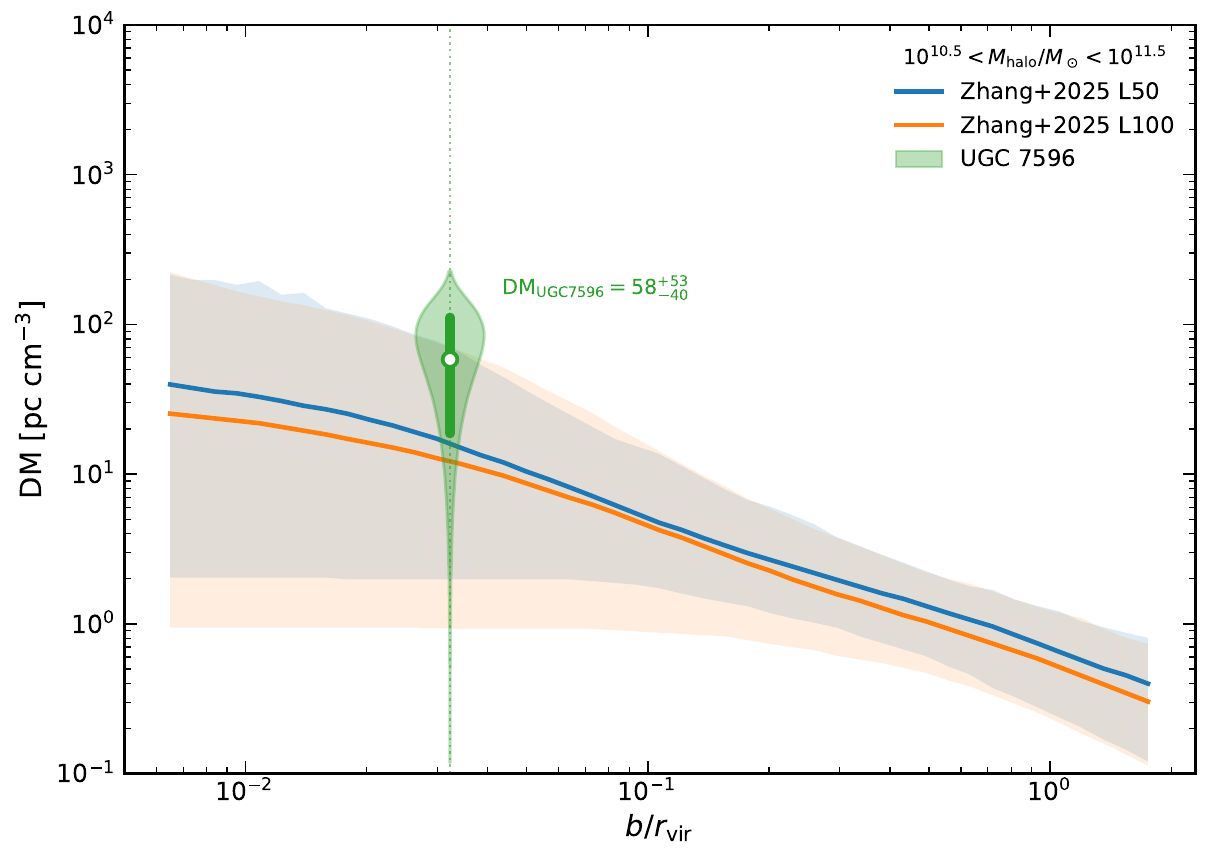}
    \caption{Comparison of the inferred UGC~7596 DM with the low mass halo ($10^{10.5}<M_{\rm halo}/M_\odot<10^{11.5}$) predictions from the CROCODILE simulations of \citet{Zhang2025}.
    The curves show the median DM profiles for the lowest halo mass bin (Figure~6 in \citet{Zhang2025}), the shaded regions show the corresponding $16$--$84$ percentile ranges, and the green point shows the inferred UGC~7596 contribution.}
    \label{fig:ugc_simulated_dm}

\end{figure*}

We also compare the inferred UGC~7596 contribution with the low mass halo DM profiles from the CROCODILE hydrodynamical simulations of \citet{Zhang2025}, adopting their lowest halo-mass bin, $10^{10.5}<M_{\rm halo}/M_\odot<10^{11.5}$.
As shown in Figure~\ref{fig:ugc_simulated_dm}, our conditional estimate, ${\rm DM}_{\rm UGC7596}=58^{+53}_{-40}~{\rm pc~cm^{-3}}$, is consistent with the broad halo-to-halo scatter predicted for halos of comparable mass.

\begin{figure*}
    \centering
    \includegraphics[width=\linewidth]{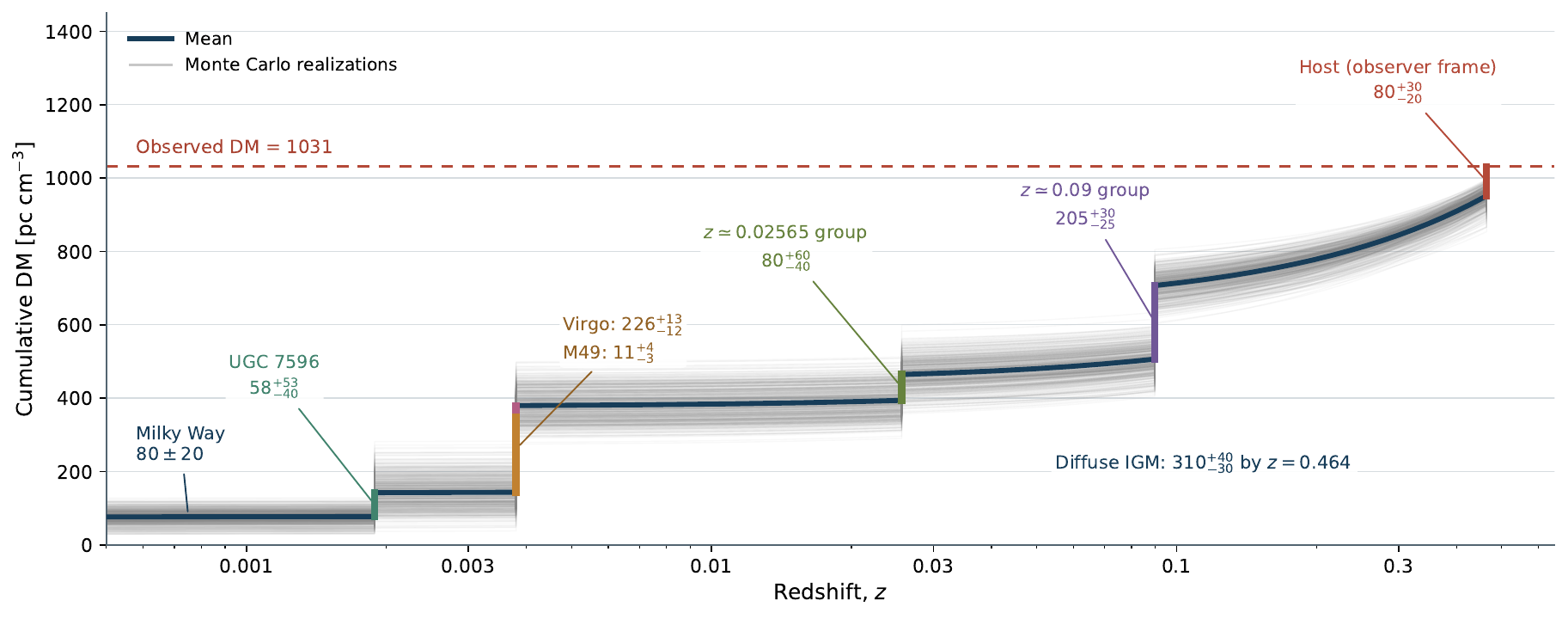}
    \caption{Cumulative DM budget along the sightline to FRB20230907D.
    Gray curves show a subset of the joint Monte Carlo realizations used in the ${\rm DM}_{\rm UGC7596}$ inference, , while the dark blue curve shows their mean cumulative DM.
    Colored vertical segments show the contributions from the MW, the UGC~7596, the Virgo cluster, the M49 group, the $z\simeq0.02565$ group, the $z\simeq0.09$ cluster, and the host galaxy.
    For each realization, the UGC~7596 contribution is the non-negative residual obtained after drawing the other terms in Equation~\ref{eq:ugc_residual}.}
    \label{fig:cumulative_dm_budget}
\end{figure*}

To visualize how the individual components contribute to the ${\rm DM}_{\rm obs}$, Figure~\ref{fig:cumulative_dm_budget} shows the cumulative DM for the Monte Carlo realizations.
For each realization, the DM contributions from the MW, the UGC~7596, the Virgo cluster, the M49 group, the $z\simeq0.02565$ group, the $z\simeq0.09$ cluster, and the host galaxy are independently drawn from their respective adopted distributions.
The UGC~7596 contribution is then calculated as the residual required by Equation~\ref{eq:ugc_residual}.
We retain only realizations for which this residual is non-negative when constructing Figure~\ref{fig:cumulative_dm_budget}.

\subsubsection{Optical disk constraint and  CGM model}
\label{sec:ugc_models}

We use the Rubin/LSST DP2 $g$-band image to constrain the contribution from the optically traced disk of UGC~7596. 
We fit its surface brightness distribution with a PSF-convolved exponential Sérsic model, $n_{\rm Sersic}=1$, using \texttt{galight} \citep{Ding2020}. 
Bad pixels, foreground stars, unrelated projected sources, and the FRB host are masked, while the central body of UGC~7596 is retained.

\begin{figure}
\centering
\includegraphics[width=\textwidth]
{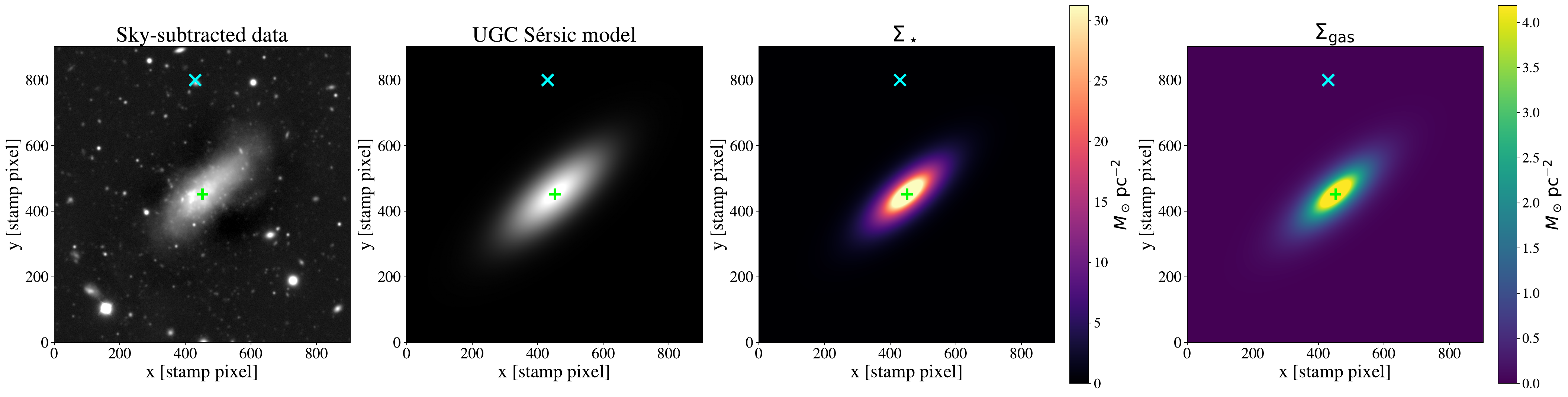}
\caption{
Rubin/LSST DP2 $g$-band modeling of UGC~7596.
From left to right, the panels show the sky subtracted image, the PSF-convolved exponential Sérsic model, the inferred stellar surface density map, and the inferred gas surface density map.
The green plus marks the fitted center of UGC~7596.
The cyan cross marks the FRB sightline.
The FRB lies slightly beyond the adopted optical radius, where the fitted disk contribution is very small.
}
\label{fig:ugc_galight_model}
\end{figure}

Figure~\ref{fig:ugc_galight_model} shows the sky subtracted image, fitted model, and inferred stellar and gas surface density maps. 
To construct the mass maps, we assume that both the stellar and gas components follow the fitted $g$-band light distribution. 
The model flux within an $80''$ aperture is normalized separately to the adopted total stellar and gas masses, and the mass assigned to each pixel is divided by its physical projected area to obtain $\Sigma_\star$ and $\Sigma_{\rm gas}$. 
These are therefore model dependent maps rather than direct resolved mass measurements.
The angular and projected separations are
\begin{equation}
\theta_{\rm FRB}=69.88'',
\qquad
b_{\rm FRB}\simeq2.64~{\rm kpc},
\qquad
\frac{\theta_{\rm FRB}}{R_{\rm opt}}\simeq1.13.
\end{equation}
The fitted S\`ersic profile is therefore evaluated slightly beyond the adopted optical radius.

\begin{figure*}
\centering
\includegraphics[width=\textwidth]
{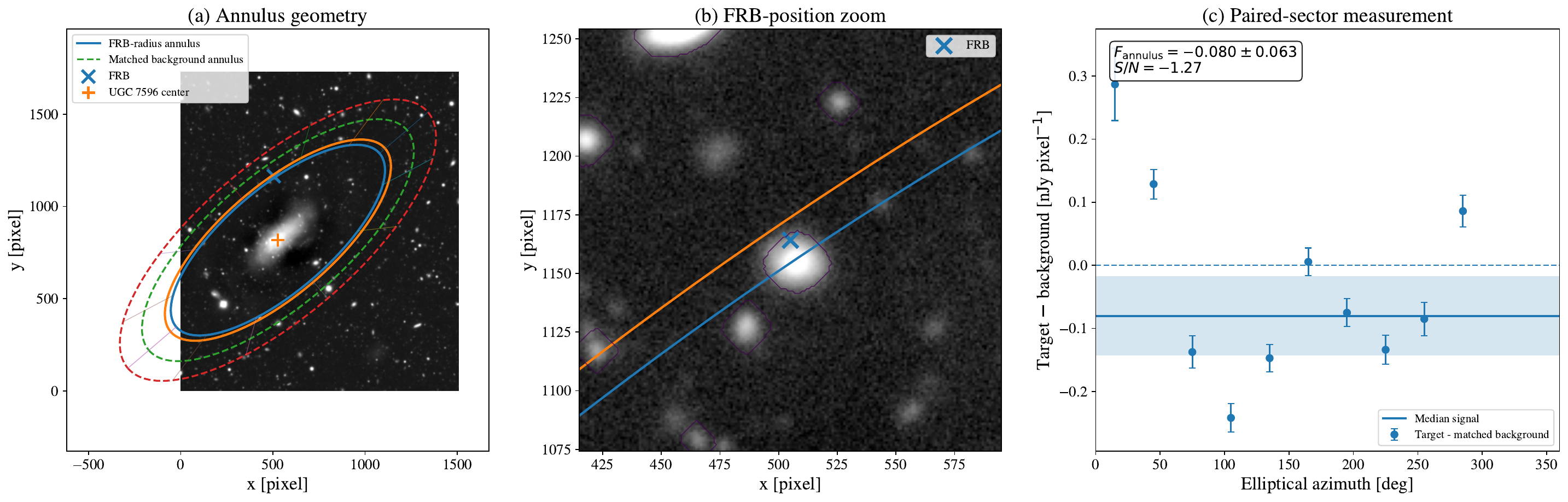}
\caption{
Diagnostic of the $g$-band surface brightness measurement at the projected FRB position around UGC~7596.
Panel (a) shows the elliptical geometry used for the measurement.
The solid curves delimit the annulus passing through the FRB radius, while the dashed curves delimit the matched background annulus.
The FRB position and the fitted center of UGC~7596 are marked.
Panel (b) shows a zoom around the FRB position and the boundaries of the FRB radius annulus.
Masked compact sources are excluded from the surface brightness measurement.
Panel (c) shows the difference between the target and matched background surface brightness for each usable azimuthal sector.
The solid horizontal line marks the median, $F_{\rm annulus}=-0.080~{\rm nJy~pixel^{-1}}$, and the shaded region shows its $1\sigma$ uncertainty.
The measurements are consistent with zero, with an overall significance of $-1.27\sigma$.
}
\label{fig:ugc_annulus_diagnostic}
\end{figure*}

To avoid interpreting the S\`ersic extrapolation at a single position as a direct detection, we also measure the image in an elliptical annulus at the FRB radius. 
The annulus is divided into azimuthal sectors, each of which is compared with a background sector at the same elliptical angle.
Compact contaminants are masked, while diffuse emission associated with UGC~7596 is retained.
Figure~\ref{fig:ugc_annulus_diagnostic} illustrates the measurement geometry, shows a zoom around the FRB position, and presents the background subtracted signal for the individual azimuthal sectors.
The median background subtracted signal is $F_{\rm annulus}=-0.080\pm0.063~{\rm nJy~pixel^{-1}}$, corresponding to a significance of $-1.27\sigma$. 
The sector measurements scatter around zero, and we therefore find no significant $g$-band emission at the FRB radius.

The corresponding conservative $3\sigma$ surface brightness limit is $\mu_g>29.71~{\rm mag~arcsec^{-2}}$.
For the ISM gas, we use the fitted optical light distribution only as a simple spatial proxy for the gas distribution.
UGC~7596 has an H\,{\sc I} mass of $M_{\rm HI}=(0.7\pm0.1)\times10^8~M_\odot$ \citep{Sorgho2017}, but its H\,{\sc I} morphology is asymmetric and offset from the optical center, and the available imaging does not resolve the gas column at the FRB sightline.
We therefore retain the optical light distribution as an illustrative model for the ISM gas.
Under the assumption that the gas follows the fitted optical light distribution, this gives $\Sigma_{\rm gas}<7.7\times10^{-3}~M_\odot~{\rm pc^{-2}}(3\sigma)$.
The corresponding ISM contribution is ${\rm DM}_{\rm ISM}<0.27\left(\frac{x_{\rm ion}}{1}\right)~{\rm pc~cm^{-3}}(3\sigma)$, where $x_{\rm ion}$ is the ionized fraction of the gas column.
Even if the entire model gas column were ionized, the contribution from the optically traced disk or ISM would be negligible compared with the other terms in the FRB20230907D DM budget.

The imaging constraint does not apply to the extended ionized CGM that is not traced by the optical light. 
For this, we apply the modified-NFW model defined in Section~\ref{sec:intervening_group_dm}, adopting
\begin{equation}
M_{\rm halo}=3\times10^{10}~M_\odot,
\qquad
f_{\rm hot}=0.75,
\qquad
r_{\rm cut}=2R_{\rm vir},
\end{equation}
with the same values of $c$, $\alpha$, and $y_0$ used for the intervening groups.
For the adopted cosmology,
\begin{equation}
R_{\rm vir}\simeq81.9~{\rm kpc},
\qquad
\frac{b_{\rm FRB}}{R_{\rm vir}}\simeq0.032.
\end{equation}
The resulting CGM contribution is
${\rm DM}_{\rm CGM,mNFW}=24^{+6}_{-5}~{\rm pc~cm^{-3}}$,
where the quoted interval propagates a $0.3$ dex scatter in halo mass while holding the profile parameters fixed.
This value is consistent with the broad raw residual distribution inferred from the Monte Carlo DM budget.

This rough estimate of the CGM assumes that UGC~7596 retains a large fraction of its cosmological baryon allotment in ionized gas and that this gas follows the same spherical profile adopted for much more massive halos. 
Uncertainties in baryon retention, $f_{\rm hot}$, concentration, and radial structure are likely larger than the quoted halo mass uncertainty.

Our residual inference and physical models  provide complementary constraints: 
the optically traced ISM contribution is negligible, whereas an extended ionized CGM could contribute several tens of ${\rm pc~cm^{-3}}$ under the adopted modified-NFW assumptions. 
This illustrative CGM value is smaller than the median of the positive conditioned residual distribution, and the raw residual remains consistent with zero.
The present analysis therefore does not require UGC~7596 to supply the full budget remainder.

\section{Conclusion}
\label{sec:conclusion}

We characterized the foreground environment of FRB20230907D using Subaru/PFS spectroscopy supplemented by SDSS spectroscopy, published low redshift group catalogs, Rubin/LSST imaging, and eROSITA X-ray data.
The objective was to identify foreground structures contributing to the large observed DM, ${\rm DM}_{\rm obs}=1031~{\rm pc~cm^{-3}}$, and to assess how these structures affect the inferred contribution associated with the nearby galaxy UGC~7596.

We constructed a combined PFS+SDSS spectroscopic catalog and applied the Nessie FoF algorithm to identify candidate foreground groups. 
Mock catalog tests showed that Nessie reliably recovers massive systems, with the best performance for the highest-mass halos.
We therefore examined the groups relevant to the FRB sightline individually. 
Within the PFS+SDSS group finding redshift range, $0.045<z<0.555$, only one recovered system extends sufficiently far to be intersected by the FRB sightline.
After removing a projected secondary component, this system lies at $z=0.089806$, contains 323 spectroscopic members, and has a velocity dispersion of approximately $776~{\rm km~s^{-1}}$. 
The velocity dispersion provides an independent estimate $M_{200,\sigma}\simeq5.2\times10^{14}~M_\odot$, while our SBI analysis gives $M_{\rm vir}=7.70\times10^{14}~M_\odot$.
Its identification is supported independently by published SDSS group catalogs and nearby eROSITA X-ray emission, while offsets among its optical and X-ray centroids suggest a dynamically complex or unrelaxed structure.

Because the PFS+SDSS group finding redshift range does not extend below $z=0.045$, we also examined published low redshift group catalogs. 
We identified an additional foreground group at $z=0.025649$, with a reported mass of approximately $1.5\times10^{14}~M_\odot$. 
This system is included as an independent component of the DM budget.

Using a common modified-NFW gas prescription with $f_{\rm hot}=0.75$, we infer observer frame contributions of $80^{+60}_{-40}~{\rm pc~cm^{-3}}$ from the $z\simeq0.02565$ group  and $205^{+30}_{-25}~{\rm pc~cm^{-3}}$ from the $z\simeq0.09$ system. 
These estimates are subject to substantial systematic uncertainties because they depend on the adopted halo masses, hot gas fractions, radial profiles, and truncation radii. 
The full budget also contains substantial contributions from the diffuse IGM and the Virgo environment, estimated as  $310^{+40}_{-30}~{\rm pc~cm^{-3}}$ and $226^{+13}_{-12}~{\rm pc~cm^{-3}}$, respectively, under the adopted models.

After subtracting the MW, the diffuse IGM, Virgo, M49, intervening group, and host contributions, while propagating their adopted uncertainties, the raw residual associated with UGC~7596 is $16^{+73}_{-86}~{\rm pc~cm^{-3}}$. 
Its $16$--$84$ percentile interval includes zero, and $57.9\%$ of the Monte Carlo realizations yield a non-negative residual. 
Conditioning on a non-negative electron column gives ${\rm DM}_{\rm UGC7596}=58^{+53}_{-40}~{\rm pc~cm^{-3}}$, but this conditional value should not be interpreted as a detection. 
The current DM budget therefore does not require a distinct positive electron column uniquely associated with UGC~7596.

Rubin/LSST imaging provides an independent constraint on the optically traced material in UGC~7596. 
We detect no significant $g$-band disk emission at the FRB position, implying ${\rm DM}_{\rm ISM}<0.27\,x_{\rm ion}~{\rm pc~cm^{-3}}$ under the assumption that the gas follows the optical light distribution. 
The optically traced disk therefore cannot account for an appreciable residual DM. 
An illustrative modified-NFW model for a $3\times10^{10}~M_\odot$ halo instead gives an extended CGM contribution of $24^{+6}_{-5}~{\rm pc~cm^{-3}}$. 
This estimate is highly model dependent because the baryon retention and gas distribution of such a low mass halo are poorly constrained.

Our results show that the high DM of FRB20230907D cannot be assigned unambiguously to a single foreground galaxy, group or cluster: several physically distinct structures contribute along the sightline, including the diffuse IGM, the Virgo environment, at least two intervening cluster scale halos, and a dwarf galaxy.
We show that dense spectroscopy is essential for identifying and modeling these structures even at relatively low redshifts, while wide area group catalogs and X-ray observations provide important complementary information. 

Among the largest uncertainties in our analysis is the modeling of identified halo masses, circum-halo gas mass fractions, as well as the diffuse cosmic web structure \citep{dong2025,dong2026}.
The FLIMFLAM survey (\citealt{Khrykin2024b},\citealt{Huang2024}) aims to address these issues through a joint analysis of FRB dispersion measures with associated foreground spectroscopic surveys.

\begin{acknowledgments}
We thank Isabel Medlock and Daisuke Nagai for useful comments.
We used ChatGPT to revise the narrative, which was then carefully fact-checked and edited.
K-GL acknowledges support from JSPS Kakenhi grant nos. JP18H05868, JP19K14755, and JP24H00241. Kavli IPMU was established by World Premier International Research Cen-ter Initiative (WPI), MEXT, Japan. This work was performed inpart at the Center for Data-Driven Discovery, Kavli IPMU (WPI), and made use of computing resources at Kavli IPMU.
I.S.K. and N.T. acknowledge support from grant ANID / FONDO ALMA 2024 / 31240053. 
I.P.M. acknowledges funding for this work from the European Research Council (ERC) under the European Union’s Horizon 2020 research and innovation programme (`EuroFlash’; Grant agreement No. 101098079). B.Y.W. gratefully acknowledge the support of the NSF-Simons AI-Institute for the Sky (SkAI) via grants NSF AST-2421845 and Simons Foundation MPS-AI-00010513.
L.T. acknowledges support from JSPS Kakenhi grants 24K22878, 26K17136, JST PRESTO JPMJPR26287514, and Royal Society International Collaboration Award Round 2 (Japan) 252140.
K.N. acknowledges support from JSPS KAKENHI grant 24H00002, 24H00241, JP25K01032, and the JSPS International Leading Research (ILR) project, JP22K21349. 
KN also acknowledges support from the Kavli IPMU, the World Premier Research Center Initiative (WPI), UTIAS, and the University of Tokyo.

The instrument \textquoteleft{}\={O}nohi\textquoteleft{}ula - Prime Focus Spectrograph (PFS) including both hardware
and software was developed by the PFS collaboration consisting of over 25 institutes
across seven countries (in alphabetical order, Brazil, China, France, Germany, Japan,
Taiwan, and the United States), where the technical activities were conducted by (in
alphabetical order) Academia Sinica Institute of Astronomy and Astrophysics (Taiwan),
California Institute of Technology, Johns Hopkins University, Kavli Institute for the
Physics and Mathematics of the Universe in the University of Tokyo (Kavli IPMU),
Laboratoire d'Astrophysique de Marseille, Laborat\`orio Nacional de Astrof\`isica (Brazil),
Max-Planck-Institut f\"ur Astrophysik, Max-Planck-Institut f\"ur extraterrestrische Physik,
NASA Jet Propulsion Laboratory, National Astronomical Observatory of Japan (NAOJ),
Princeton University, and Universidade de S\~ao Paulo under the oversight by Project
Office hosted by Kavli IPMU (later NAOJ). There were also essential commitments from
academic and industrial partners such as Durham University (United Kingdom) and
Bertin Technologies (France).

This work is based (in part) on data collected at the Subaru Telescope, which is
operated by the National Astronomical Observatory of Japan. We are honored and
grateful for the opportunity of observing the Universe from Maunakea, which has the
cultural, historical, and natural significance in Hawaii.

We appreciate the development and operation of PFS Science Platform by
Subaru Telescope and Astronomy Data Center at NAOJ which enables access to
both PFS and HSC data and various analyses on the server side.

This material is based upon work supported in part by the National Science Foundation through Cooperative Agreements AST-1258333 and AST-2241526 and Cooperative Support Agreements AST-1202910 and 2211468 managed by the Association of Universities for Research in Astronomy (AURA), and the Department of Energy under Contract No. DE-AC02-76SF00515 with the SLAC National Accelerator Laboratory managed by Stanford University. Additional Rubin Observatory funding comes from private donations, grants to universities, and in-kind support from LSST-DA Institutional Members.

This publication is based in part on proprietary Rubin Observatory Legacy Survey of Space and Time (LSST) data, and was prepared in accordance with the Rubin Observatory data rights and access policies. All authors of this publication meet the requirements for co-authorship of proprietary LSST data.

This research uses services or data provided by the Rubin Science Platform at NSF-DOE Vera C. Rubin Observatory, which is jointly funded by the U.S. National Science Foundation and the U.S. Department of Energy, Office of Science.
\end{acknowledgments}

\appendix
\section{Simulation-Based Inference Setup}
\label{app:sbi}

Our simulation-based inference (SBI) framework for cluster mass estimation follows \cite{Wang2026}, with models retrained on Uchuu UniverseMachine~\citep{Ishiyama2021,Aung2023} clusters at $z\simeq0.09$ to match the massive foreground cluster at that redshift observed by PFS. The framework combines interloper identification with cluster mass posterior inference. We construct mock cylindrical catalogs using projected spatial and line-of-sight velocity selections, naturally including cluster members and foreground/background interlopers. Below, we describe the modifications introduced to mimic PFS observational incompleteness and present the resulting interloper identification and mass inference.

\begin{figure}[!htb]
    \centering
    \includegraphics[width=0.99\linewidth]{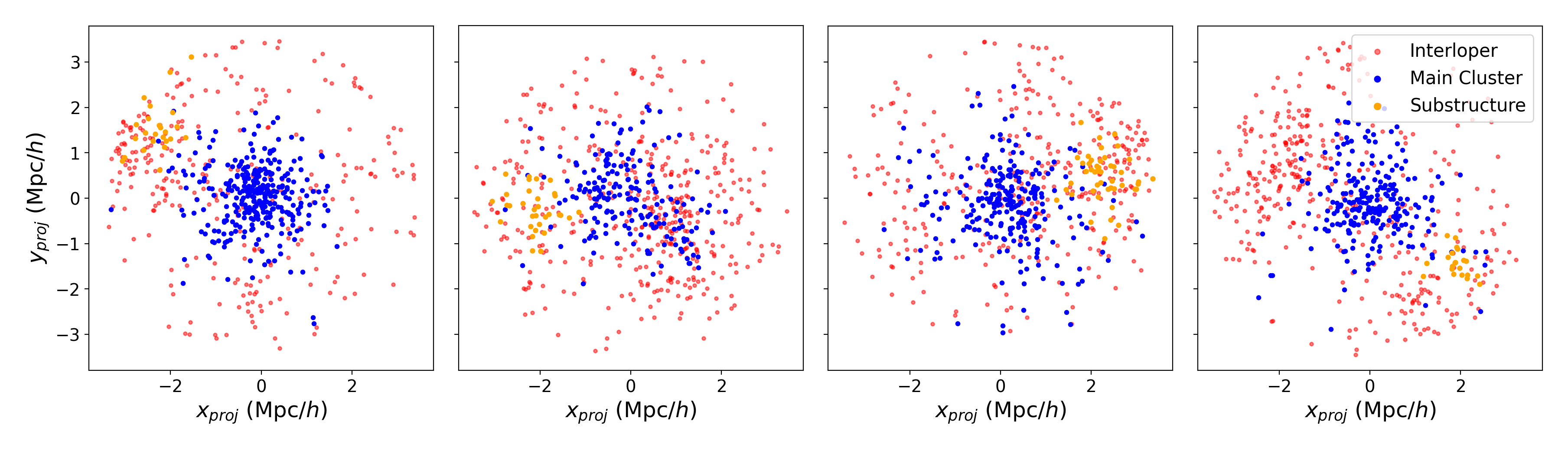}
    \caption{Examples of four Uchuu galaxy clusters at $z\simeq0.09$ exhibiting prominent secondary galaxy concentrations. Each panel shows the projected spatial distribution of galaxies in a cylindrical catalog centered on the main cluster. Blue points represent main-cluster members, orange points indicate galaxies associated with a secondary substructure, and red points denote foreground/background interlopers.}
\label{fig:uchuu_substructure}
\end{figure}

First, to ensure that the Uchuu training sample captures the two prominent galaxy concentrations observed in the PFS foreground cluster as mentioned in Section~\ref{sec:z009_system}, we search for analogous configurations in the simulation. Figure~\ref{fig:uchuu_substructure} shows four sample Uchuu systems exhibiting similar main-cluster and substructure configurations, demonstrating that such systems are represented in the training data.

\begin{figure}
    \centering
    \includegraphics[width=0.999\linewidth]{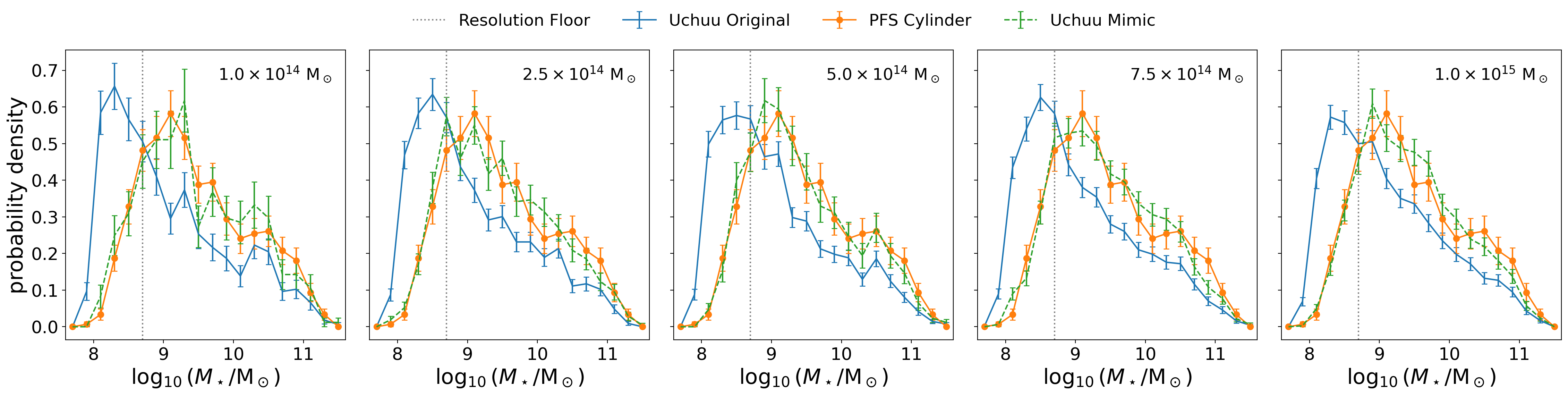}
    \caption{
    Comparison of the stellar-mass distributions of the original Uchuu cylindrical catalogs (blue), the observed PFS cylindrical sample (orange), and the PFS-mimicked Uchuu catalogs (green dashed) for five clusters spanning different halo masses. The vertical dotted line indicates the adopted Uchuu stellar-mass resolution limit. Error bars represent Poisson uncertainties in the estimated probability densities.
    }
    \label{fig:pfs_completeness}
\end{figure}

To account for the stellar-mass-dependent incompleteness of the PFS observations, we introduce a sigmoid selection function calibrated to the stellar-mass distribution of the observed PFS cylindrical catalog. Each Uchuu galaxy is assigned a retention probability that increases with stellar mass, preferentially excluding low-mass galaxies while preserving a larger fraction of massive systems. We additionally impose a lower stellar-mass threshold of $5\times10^{8}\,{\rm M}_{\odot}$, corresponding to the adopted Uchuu stellar-mass resolution limit. Figure~\ref{fig:pfs_completeness} compares the stellar-mass distributions of the original Uchuu cylindrical catalogs, the observed PFS sample, and the PFS-mimicked Uchuu catalogs for five randomly selected clusters spanning a range of halo masses. The incompleteness prescription substantially improves the agreement between the mock and observed stellar-mass distributions, providing a more representative training sample for subsequent SBI analyses.

Moreover, to reproduce the observed galaxy richness, we apply an additional mass-independent random downsampling when the number of selected mock galaxies exceeds that of the PFS cylindrical catalog ($\sim632$ galaxies). This procedure limits the number of galaxies in each mock cylinder while retaining the stellar-mass dependence introduced by the sigmoid selection function. Finally, we also add Gaussian random noise with a standard deviation of $50\,\mathrm{km\,s^{-1}}$ to the line-of-sight velocities of the mock galaxies to mimic observational velocity uncertainties.

\begin{figure}[!b]
    \centering
    \includegraphics[width=0.35\linewidth]{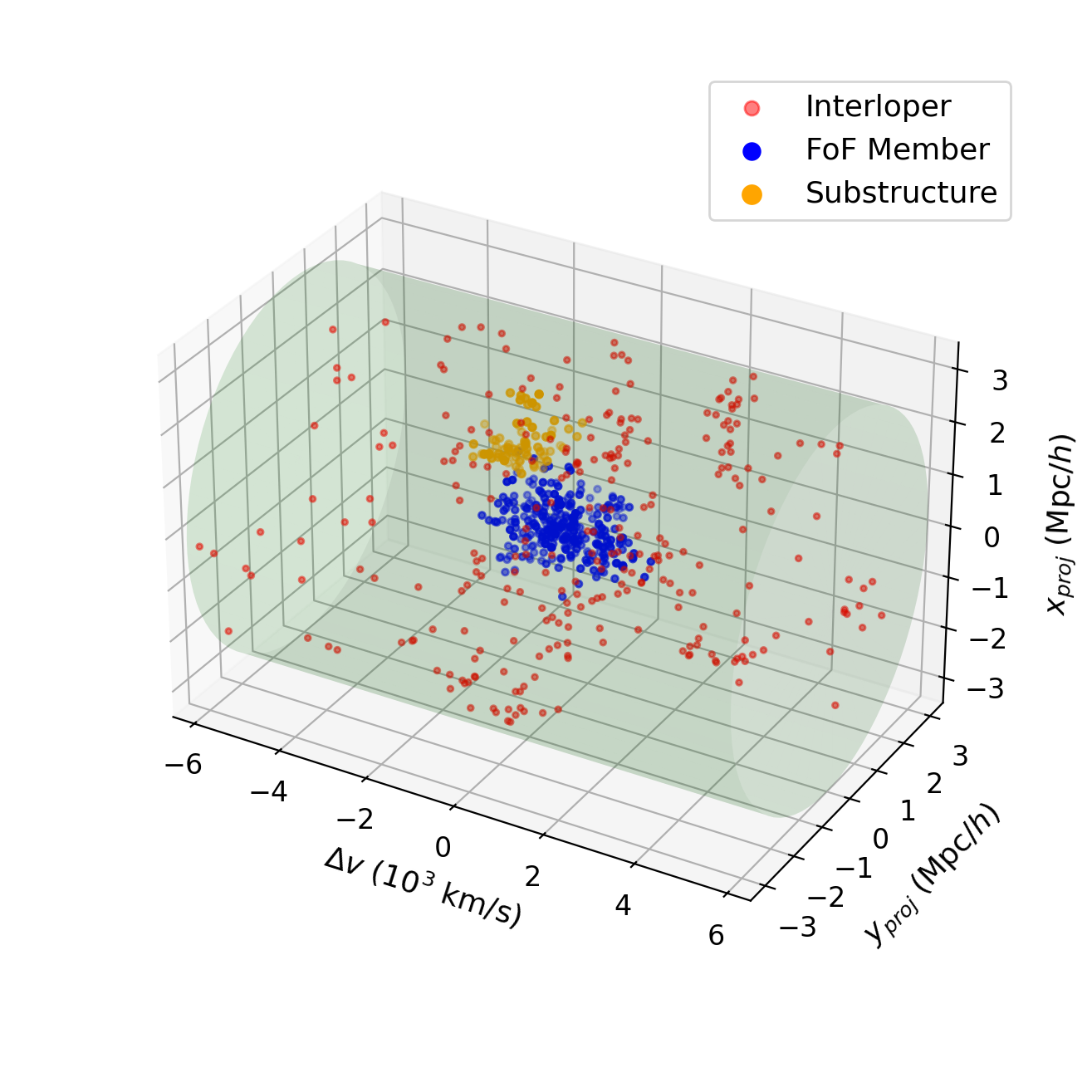}
    \includegraphics[width=0.64\linewidth]{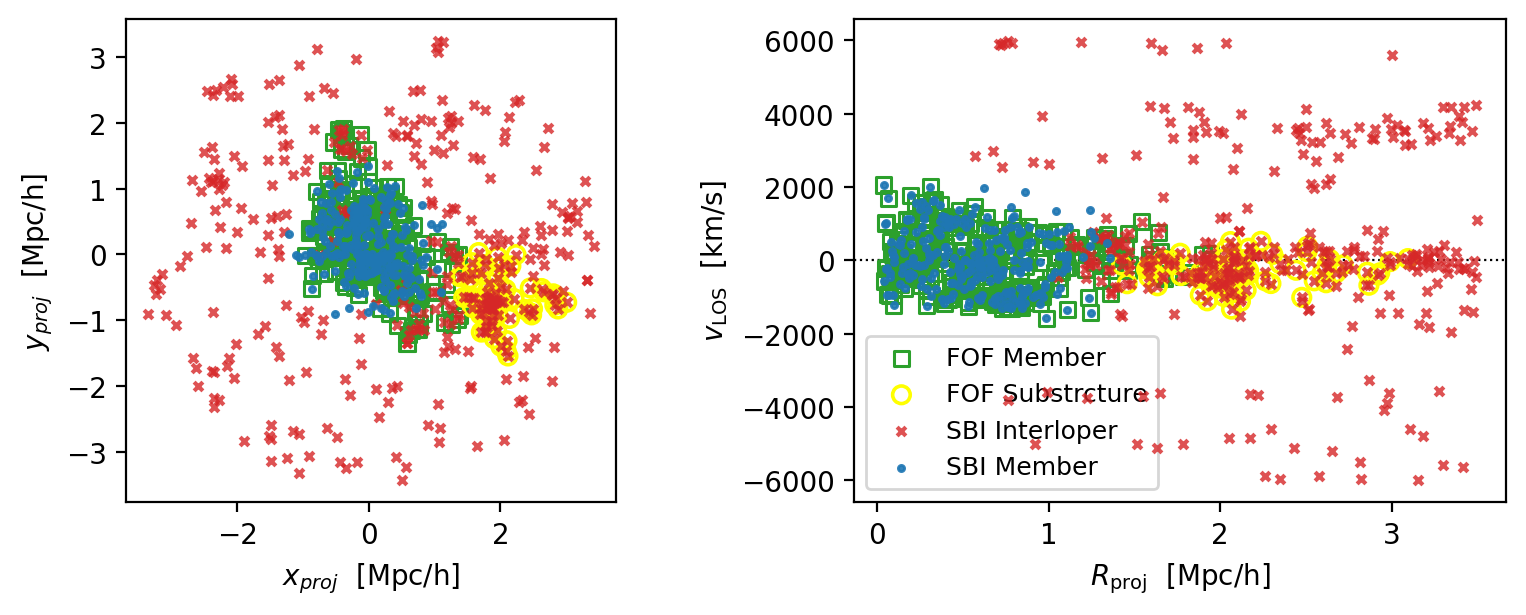}
    \caption{
    Illustration of the cylindrical galaxy selection and SBI-based interloper identification for the massive foreground cluster at $z\simeq0.09$.
    Left: Three-dimensional distribution of galaxies within the cylindrical selection volume, distinguishing Nessie FoF members (blue), substructures (orange), and interlopers (red).
    Middle: Projected spatial distribution of galaxies, highlighting the main FoF cluster and associated substructures against the surrounding field population.
    Right: Projected phase-space distribution comparing SBI-identified members (blue dots) and interlopers (red crosses) with FoF-identified members (green squares) and substructures (yellow circles).
    }
    \label{fig:cylinder}
\end{figure}

We apply the interloper identification model developed in \cite{Wang2026} to distinguish cluster members from foreground and background contaminants. The model uses the projected positions and line-of-sight velocities of galaxies to estimate their membership probabilities, thereby providing a probabilistic alternative to membership assignments based on the friends-of-friends (FoF) halo finder.

Figure~\ref{fig:cylinder} illustrates the cylindrical selection and the resulting membership classification. The three-dimensional and projected distributions reveal that the observational field cylinder contains not only the main cluster but also nearby substructures and unrelated field galaxies. The projected phase-space diagram on the right further compares the SBI classifications with the FoF-based assignments. Notably, the SBI identifier separates the dominant cluster population from a distinct substructure that is originally included within the same FoF group. 


\begin{table} [!htb]
    \centering
    \caption{Spatial and line-of-sight velocity offsets applied to the
    adopted cluster center for three random realizations used to test
    the robustness of cluster mass inference.}
    \label{tab:center_perturbations}
    \begin{tabular}{lccc}
        \hline
        & Random 1 & Random 2 & Random 3 \\
        \hline
        $\Delta x$ ($h^{-1}\,\mathrm{Mpc}$)
            & +0.06 & 0 & +0.03 \\
        $\Delta y$ ($h^{-1}\,\mathrm{Mpc}$)
            & 0 & +0.08 & +0.03 \\
        $\Delta v_{\mathrm{los}}$ ($\mathrm{km\,s^{-1}}$)
            & +60 & -90 & +60 \\
        \hline
    \end{tabular}
\end{table}

\begin{figure}[!htb]
    \centering
    \includegraphics[width=0.6\linewidth]{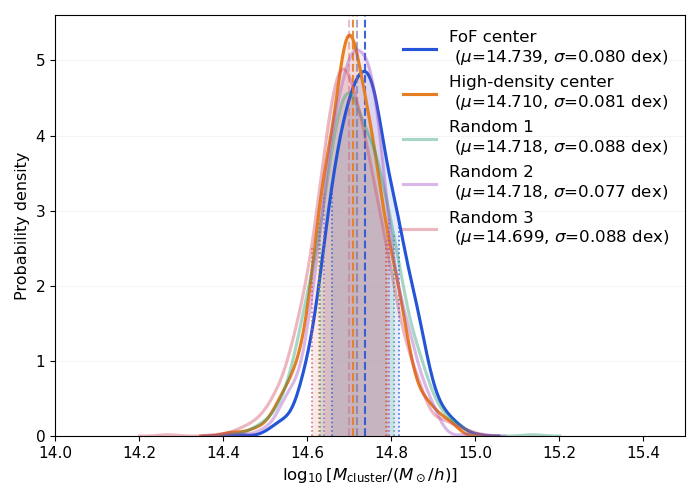}
    \caption{Posterior distributions of the inferred cluster mass under different definitions of the cluster center. The blue and orange curves correspond to the FoF and high-density centers, respectively, while the remaining curves represent three randomly perturbed centers (Table~\ref{tab:center_perturbations}). Dashed vertical lines indicate the posterior means, and dotted lines mark the corresponding $1\sigma$ intervals.}
    \label{fig:mass_posterior}
\end{figure}

Following interloper identification, we apply the mass inference framework of \cite{Wang2026} to the PFS-mimicked cylindrical catalogs. The mass inference model is retrained using the incompleteness-matched Uchuu catalogs described above, ensuring that the training data reproduce the observational selection effects of the PFS sample. We then apply the trained model to the observed PFS galaxy population to obtain the cluster mass posterior. We note that our model is trained assuming that the cluster center is accurately identified, whereas the center of the observed cluster may be subject to positional and velocity uncertainties. Therefore, to assess the robustness of the mass inference to the adopted cluster center, we compare the FOF and high-density centers as well as the three additional centers perturbed in projected position and line-of-sight velocity (Table~\ref{tab:center_perturbations}).

Figure~\ref{fig:mass_posterior} shows that all five center definitions yield consistent mass posteriors, with mean inferred masses ranging from $10^{14.699}$ to $10^{14.739}$ M$_{\odot}$/$h$ and statistical uncertainties of $0.077$--$0.088$ dex. The maximum difference between posterior means is only $0.040$ dex, smaller than the typical $1\sigma$ uncertainty of $\sim0.08$ dex. This agreement indicates that the mass inference is robust to the tested variations in the cluster center, suggesting that reasonable uncertainties in the adopted center do not significantly affect the inferred cluster mass. This test further provides confidence in the robustness of our mass estimates using the SBI pipeline.

\bibliographystyle{aasjournalv7}
\bibliography{sample701}



\end{document}